\documentstyle[mncite]{mn}
\def\PsfigVersion{1.9}
\ifx\undefined\psfig\else\endinput\fi

\let\LaTeXAtSign=\@
\let\@=\relax
\edef\psfigRestoreAt{\catcode`\@=\number\catcode`@\relax}
\catcode`\@=11\relax
\newwrite\@unused
\def\ps@typeout#1{{\let\protect\string\immediate\write\@unused{#1}}}
\ps@typeout{psfig/tex \PsfigVersion}

\def\figurepath{./}

\def\@nnil{\@nil}
\def\@empty{}
\def\@psdonoop#1\@@#2#3{}
\def\@psdo#1:=#2\do#3{\edef\@psdotmp{#2}\ifx\@psdotmp\@empty \else
\expandafter\@psdoloop#2,\@nil,\@nil\@@#1{#3}\fi}
\def\@psdoloop#1,#2,#3\@@#4#5{\def#4{#1}\ifx #4\@nnil \else
#5\def#4{#2}\ifx #4\@nnil \else#5\@ipsdoloop #3\@@#4{#5}\fi\fi}
\def\@ipsdoloop#1,#2\@@#3#4{\def#3{#1}\ifx #3\@nnil
\let\@nextwhile=\@psdonoop \else
#4\relax\let\@nextwhile=\@ipsdoloop\fi\@nextwhile#2\@@#3{#4}}
\def\@tpsdo#1:=#2\do#3{\xdef\@psdotmp{#2}\ifx\@psdotmp\@empty \else
\@tpsdoloop#2\@nil\@nil\@@#1{#3}\fi}
\def\@tpsdoloop#1#2\@@#3#4{\def#3{#1}\ifx #3\@nnil
\let\@nextwhile=\@psdonoop \else
#4\relax\let\@nextwhile=\@tpsdoloop\fi\@nextwhile#2\@@#3{#4}}
\ifx\undefined\fbox
\newdimen\fboxrule
\newdimen\fboxsep
\newdimen\ps@tempdima
\newbox\ps@tempboxa
\long\def\fbox#1{\leavevmode\setbox\ps@tempboxa\hbox{#1}\ps@tempdima\fboxrule
\advance\ps@tempdima \fboxsep \advance\ps@tempdima \dp\ps@tempboxa
\hbox{\lower \ps@tempdima\hbox
{\vbox{\hrule height \fboxrule
\hbox{\vrule width \fboxrule \hskip\fboxsep
\vbox{\vskip\fboxsep \box\ps@tempboxa\vskip\fboxsep}\hskip
\fboxsep\vrule width \fboxrule}
\hrule height \fboxrule}}}}
\fi
\newread\ps@stream
\newif\ifnot@eof       % continue looking for the bounding box?
\newif\if@noisy        % report what you're making?
\newif\if@atend        % %%BoundingBox: has (at end) specification
\newif\if@psfile       % does this look like a PostScript file?
{\catcode`\%=12\global\gdef\epsf@start{%!}}
\def\epsf@PS{PS}
\def\epsf@getbb#1{%
\openin\ps@stream=#1
\ifeof\ps@stream\ps@typeout{Error, File #1 not found}\else
{\not@eoftrue \chardef\other=12
\def\do##1{\catcode`##1=\other}\dospecials \catcode`\ =10
\loop
\if@psfile
	  \read\ps@stream to \epsf@fileline
\else{
	  \obeyspaces
\read\ps@stream to \epsf@tmp\global\let\epsf@fileline\epsf@tmp}
\fi
\ifeof\ps@stream\not@eoffalse\else
\if@psfile\else
\expandafter\epsf@test\epsf@fileline:. \\%
\fi
\expandafter\epsf@aux\epsf@fileline:. \\%
\fi
\ifnot@eof\repeat
}\closein\ps@stream\fi}%
\long\def\epsf@test#1#2#3:#4\\{\def\epsf@testit{#1#2}
			\ifx\epsf@testit\epsf@start\else
\ps@typeout{Warning! File does not start with `\epsf@start'.  It may not be a PostScript file.}
			\fi
			\@psfiletrue} % don't test after 1st line
{\catcode`\%=12\global\let\epsf@percent=%\global\def\epsf@bblit{%BoundingBox}}
\long\def\epsf@aux#1#2:#3\\{\ifx#1\epsf@percent
\def\epsf@testit{#2}\ifx\epsf@testit\epsf@bblit
	\@atendfalse
\epsf@atend #3 . \\%
	\if@atend	
	   \if@verbose{
		\ps@typeout{psfig: found `(atend)'; continuing search}
	   }\fi
\else
\epsf@grab #3 . . . \\%
\not@eoffalse
\global\no@bbfalse
\fi
\fi\fi}%
\def\epsf@grab #1 #2 #3 #4 #5\\{%
\global\def\epsf@llx{#1}\ifx\epsf@llx\empty
\epsf@grab #2 #3 #4 #5 .\\\else
\global\def\epsf@lly{#2}%
\global\def\epsf@urx{#3}\global\def\epsf@ury{#4}\fi}%
\def\epsf@atendlit{(atend)}
\def\epsf@atend #1 #2 #3\\{%
\def\epsf@tmp{#1}\ifx\epsf@tmp\empty
\epsf@atend #2 #3 .\\\else
\ifx\epsf@tmp\epsf@atendlit\@atendtrue\fi\fi}

\chardef\psletter = 11 % won't conflict with \begin{letter} now...
\chardef\other = 12

\newif \ifdebug %%% turn me on to see TeX hard at work ...
\newif\ifc@mpute %%% don't need to compute some values
\c@mputetrue % but assume that we do

\let\then = \relax
\def\r@dian{pt }
\let\r@dians = \r@dian
\let\dimensionless@nit = \r@dian
\let\dimensionless@nits = \dimensionless@nit
\def\internal@nit{sp }
\let\internal@nits = \internal@nit
\newif\ifstillc@nverging
\def \Mess@ge #1{\ifdebug \then \message {#1} \fi}

{ %%% Things that need abnormal catcodes %%%
	\catcode `\@ = \psletter
	\gdef \nodimen {\expandafter \n@dimen \the \dimen}
	\gdef \term #1 #2 #3%
	       {\edef \t@ {\the #1}%%% freeze parameter 1 (count, by value)
		\edef \t@@ {\expandafter \n@dimen \the #2\r@dian}%
		\t@rm {\t@} {\t@@} {#3}%
	       }
	\gdef \t@rm #1 #2 #3%
	       {{%
		\count 0 = 0
		\dimen 0 = 1 \dimensionless@nit
		\dimen 2 = #2\relax
		\Mess@ge {Calculating term #1 of \nodimen 2}%
		\loop
		\ifnum	\count 0 < #1
		\then	\advance \count 0 by 1
			\Mess@ge {Iteration \the \count 0 \space}%
			\Multiply \dimen 0 by {\dimen 2}%
			\Mess@ge {After multiplication, term = \nodimen 0}%
			\Divide \dimen 0 by {\count 0}%
			\Mess@ge {After division, term = \nodimen 0}%
		\repeat
		\Mess@ge {Final value for term #1 of
				\nodimen 2 \space is \nodimen 0}%
		\xdef \Term {#3 = \nodimen 0 \r@dians}%
		\aftergroup \Term
	       }}
	\catcode `\p = \other
	\catcode `\t = \other
	\gdef \n@dimen #1pt{#1} %%% throw away the ``pt''
}

\def \Divide #1by #2{\divide #1 by #2} %%% just a synonym

\def \Multiply #1by #2%%% allows division of a dimen by a dimen
{{%%% should really freeze parameter 2 (dimen, passed by value)
	\count 0 = #1\relax
	\count 2 = #2\relax
	\count 4 = 65536
	\Mess@ge {Before scaling, count 0 = \the \count 0 \space and
			count 2 = \the \count 2}%
	\ifnum	\count 0 > 32767 %%% do our best to avoid overflow
	\then	\divide \count 0 by 4
		\divide \count 4 by 4
	\else	\ifnum	\count 0 < -32767
		\then	\divide \count 0 by 4
			\divide \count 4 by 4
		\else
		\fi
	\fi
	\ifnum	\count 2 > 32767 %%% while retaining reasonable accuracy
	\then	\divide \count 2 by 4
		\divide \count 4 by 4
	\else	\ifnum	\count 2 < -32767
		\then	\divide \count 2 by 4
			\divide \count 4 by 4
		\else
		\fi
	\fi
	\multiply \count 0 by \count 2
	\divide \count 0 by \count 4
	\xdef \product {#1 = \the \count 0 \internal@nits}%
	\aftergroup \product
}}

\def\r@duce{\ifdim\dimen0 > 90\r@dian \then   % sin(x+90) = sin(180-x)
		\multiply\dimen0 by -1
		\advance\dimen0 by 180\r@dian
		\r@duce
	    \else \ifdim\dimen0 < -90\r@dian \then  % sin(-x) = sin(360+x)
		\advance\dimen0 by 360\r@dian
		\r@duce
		\fi
	    \fi}

\def\Sine#1%
{{%
	\dimen 0 = #1 \r@dian
	\r@duce
	\ifdim\dimen0 = -90\r@dian \then
	   \dimen4 = -1\r@dian
	   \c@mputefalse
	\fi
	\ifdim\dimen0 = 90\r@dian \then
	   \dimen4 = 1\r@dian
	   \c@mputefalse
	\fi
	\ifdim\dimen0 = 0\r@dian \then
	   \dimen4 = 0\r@dian
	   \c@mputefalse
	\fi
	\ifc@mpute \then
		\divide\dimen0 by 180
		\dimen0=3.141592654\dimen0
		\dimen 2 = 3.1415926535897963\r@dian %%% a well-known constant
		\divide\dimen 2 by 2 %%% we only deal with -pi/2 : pi/2
		\Mess@ge {Sin: calculating Sin of \nodimen 0}%
		\count 0 = 1 %%% see power-series expansion for sine
		\dimen 2 = 1 \r@dian %%% ditto
		\dimen 4 = 0 \r@dian %%% ditto
		\loop
			\ifnum	\dimen 2 = 0 %%% then we've done
			\then	\stillc@nvergingfalse
			\else	\stillc@nvergingtrue
			\fi
			\ifstillc@nverging %%% then calculate next term
			\then	\term {\count 0} {\dimen 0} {\dimen 2}%
				\advance \count 0 by 2
				\count 2 = \count 0
				\divide \count 2 by 2
				\ifodd	\count 2 %%% signs alternate
				\then	\advance \dimen 4 by \dimen 2
				\else	\advance \dimen 4 by -\dimen 2
				\fi
		\repeat
	\fi		
			\xdef \sine {\nodimen 4}%
}}

\def\Cosine#1{\ifx\sine\UnDefined\edef\Savesine{\relax}\else
		             \edef\Savesine{\sine}\fi
	{\dimen0=#1\r@dian\advance\dimen0 by 90\r@dian
	 \Sine{\nodimen 0}
	 \xdef\cosine{\sine}
	 \xdef\sine{\Savesine}}}	
\def\psdraft{
	\def\@psdraft{0}
}
\def\psfull{
	\def\@psdraft{100}
}

\psfull

\newif\if@scalefirst
\def\psscalefirst{\@scalefirsttrue}
\def\psrotatefirst{\@scalefirstfalse}
\psrotatefirst

\newif\if@draftbox
\def\psnodraftbox{
	\@draftboxfalse
}
\def\psdraftbox{
	\@draftboxtrue
}
\@draftboxtrue

\newif\if@prologfile
\newif\if@postlogfile
\def\pssilent{
	\@noisyfalse
}
\def\psnoisy{
	\@noisytrue
}
\psnoisy
\newif\if@bbllx
\newif\if@bblly
\newif\if@bburx
\newif\if@bbury
\newif\if@height
\newif\if@width
\newif\if@rheight
\newif\if@rwidth
\newif\if@angle
\newif\if@clip
\newif\if@verbose
\def\@p@@sclip#1{\@cliptrue}

\newif\if@decmpr

\def\@p@@sfigure#1{\def\@p@sfile{null}\def\@p@sbbfile{null}
	        \openin1=#1.bb
		\ifeof1\closein1
	        	\openin1=\figurepath#1.bb
			\ifeof1\closein1
			        \openin1=#1
				\ifeof1\closein1%
				       \openin1=\figurepath#1
					\ifeof1
					   \ps@typeout{Error, File #1 not found}
						\if@bbllx\if@bblly
				   		\if@bburx\if@bbury
			      				\def\@p@sfile{#1}%
			      				\def\@p@sbbfile{#1}%
							\@decmprfalse
				  	   	\fi\fi\fi\fi
					\else\closein1
				    		\def\@p@sfile{\figurepath#1}%
				    		\def\@p@sbbfile{\figurepath#1}%
						\@decmprfalse
	                       		\fi%
			 	\else\closein1%
					\def\@p@sfile{#1}
					\def\@p@sbbfile{#1}
					\@decmprfalse
			 	\fi
			\else
				\def\@p@sfile{\figurepath#1}
				\def\@p@sbbfile{\figurepath#1.bb}
				\@decmprtrue
			\fi
		\else
			\def\@p@sfile{#1}
			\def\@p@sbbfile{#1.bb}
			\@decmprtrue
		\fi}

\def\@p@@sfile#1{\@p@@sfigure{#1}}

\def\@p@@sbbllx#1{
		\@bbllxtrue
		\dimen100=#1
		\edef\@p@sbbllx{\number\dimen100}
}
\def\@p@@sbblly#1{
		\@bbllytrue
		\dimen100=#1
		\edef\@p@sbblly{\number\dimen100}
}
\def\@p@@sbburx#1{
		\@bburxtrue
		\dimen100=#1
		\edef\@p@sbburx{\number\dimen100}
}
\def\@p@@sbbury#1{
		\@bburytrue
		\dimen100=#1
		\edef\@p@sbbury{\number\dimen100}
}
\def\@p@@sheight#1{
		\@heighttrue
		\dimen100=#1
		\edef\@p@sheight{\number\dimen100}
}
\def\@p@@swidth#1{
		\@widthtrue
		\dimen100=#1
		\edef\@p@swidth{\number\dimen100}
}
\def\@p@@srheight#1{
		\@rheighttrue
		\dimen100=#1
		\edef\@p@srheight{\number\dimen100}
}
\def\@p@@srwidth#1{
		\@rwidthtrue
		\dimen100=#1
		\edef\@p@srwidth{\number\dimen100}
}
\def\@p@@sangle#1{
		\@angletrue
		\edef\@p@sangle{#1} %\number\dimen100}
}
\def\@p@@ssilent#1{
		\@verbosefalse
}
\def\@p@@sprolog#1{\@prologfiletrue\def\@prologfileval{#1}}
\def\@p@@spostlog#1{\@postlogfiletrue\def\@postlogfileval{#1}}
\def\@cs@name#1{\csname #1\endcsname}
\def\@setparms#1=#2,{\@cs@name{@p@@s#1}{#2}}
\def\ps@init@parms{
		\@bbllxfalse \@bbllyfalse
		\@bburxfalse \@bburyfalse
		\@heightfalse \@widthfalse
		\@rheightfalse \@rwidthfalse
		\def\@p@sbbllx{}\def\@p@sbblly{}
		\def\@p@sbburx{}\def\@p@sbbury{}
		\def\@p@sheight{}\def\@p@swidth{}
		\def\@p@srheight{}\def\@p@srwidth{}
		\def\@p@sangle{0}
		\def\@p@sfile{} \def\@p@sbbfile{}
		\def\@p@scost{10}
		\def\@sc{}
		\@prologfilefalse
		\@postlogfilefalse
		\@clipfalse
		\if@noisy
			\@verbosetrue
		\else
			\@verbosefalse
		\fi
}
\def\parse@ps@parms#1{
	 	\@psdo\@psfiga:=#1\do
		   {\expandafter\@setparms\@psfiga,}}
\newif\ifno@bb
\def\bb@missing{
	\if@verbose{
		\ps@typeout{psfig: searching \@p@sbbfile \space  for bounding box}
	}\fi
	\no@bbtrue
	\epsf@getbb{\@p@sbbfile}
\ifno@bb \else \bb@cull\epsf@llx\epsf@lly\epsf@urx\epsf@ury\fi
}	
\def\bb@cull#1#2#3#4{
	\dimen100=#1 bp\edef\@p@sbbllx{\number\dimen100}
	\dimen100=#2 bp\edef\@p@sbblly{\number\dimen100}
	\dimen100=#3 bp\edef\@p@sbburx{\number\dimen100}
	\dimen100=#4 bp\edef\@p@sbbury{\number\dimen100}
	\no@bbfalse
}
\newdimen\p@intvaluex
\newdimen\p@intvaluey
\def\rotate@#1#2{{\dimen0=#1 sp\dimen1=#2 sp
		  \global\p@intvaluex=\cosine\dimen0
		  \dimen3=\sine\dimen1
		  \global\advance\p@intvaluex by -\dimen3
		  \global\p@intvaluey=\sine\dimen0
		  \dimen3=\cosine\dimen1
		  \global\advance\p@intvaluey by \dimen3
		  }}
\def\compute@bb{
		\no@bbfalse
		\if@bbllx \else \no@bbtrue \fi
		\if@bblly \else \no@bbtrue \fi
		\if@bburx \else \no@bbtrue \fi
		\if@bbury \else \no@bbtrue \fi
		\ifno@bb \bb@missing \fi
		\ifno@bb \ps@typeout{FATAL ERROR: no bb supplied or found}
			\no-bb-error
		\fi
		\count203=\@p@sbburx
		\count204=\@p@sbbury
		\advance\count203 by -\@p@sbbllx
		\advance\count204 by -\@p@sbblly
		\edef\ps@bbw{\number\count203}
		\edef\ps@bbh{\number\count204}
		\if@angle
			\Sine{\@p@sangle}\Cosine{\@p@sangle}
	        	{\dimen100=\maxdimen\xdef\r@p@sbbllx{\number\dimen100}
					    \xdef\r@p@sbblly{\number\dimen100}
			                    \xdef\r@p@sbburx{-\number\dimen100}
					    \xdef\r@p@sbbury{-\number\dimen100}}
\def\minmaxtest{
			   \ifnum\number\p@intvaluex<\r@p@sbbllx
			      \xdef\r@p@sbbllx{\number\p@intvaluex}\fi
			   \ifnum\number\p@intvaluex>\r@p@sbburx
			      \xdef\r@p@sbburx{\number\p@intvaluex}\fi
			   \ifnum\number\p@intvaluey<\r@p@sbblly
			      \xdef\r@p@sbblly{\number\p@intvaluey}\fi
			   \ifnum\number\p@intvaluey>\r@p@sbbury
			      \xdef\r@p@sbbury{\number\p@intvaluey}\fi
			   }
			\rotate@{\@p@sbbllx}{\@p@sbblly}
			\minmaxtest
			\rotate@{\@p@sbbllx}{\@p@sbbury}
			\minmaxtest
			\rotate@{\@p@sbburx}{\@p@sbblly}
			\minmaxtest
			\rotate@{\@p@sbburx}{\@p@sbbury}
			\minmaxtest
			\edef\@p@sbbllx{\r@p@sbbllx}\edef\@p@sbblly{\r@p@sbblly}
			\edef\@p@sbburx{\r@p@sbburx}\edef\@p@sbbury{\r@p@sbbury}
		\fi
		\count203=\@p@sbburx
		\count204=\@p@sbbury
		\advance\count203 by -\@p@sbbllx
		\advance\count204 by -\@p@sbblly
		\edef\@bbw{\number\count203}
		\edef\@bbh{\number\count204}
}
\def\in@hundreds#1#2#3{\count240=#2 \count241=#3
		     \count100=\count240	% 100 is first digit #2/#3
		     \divide\count100 by \count241
		     \count101=\count100
		     \multiply\count101 by \count241
		     \advance\count240 by -\count101
		     \multiply\count240 by 10
		     \count101=\count240	%101 is second digit of #2/#3
		     \divide\count101 by \count241
		     \count102=\count101
		     \multiply\count102 by \count241
		     \advance\count240 by -\count102
		     \multiply\count240 by 10
		     \count102=\count240	% 102 is the third digit
		     \divide\count102 by \count241
		     \count200=#1\count205=0
		     \count201=\count200
			\multiply\count201 by \count100
		 	\advance\count205 by \count201
		     \count201=\count200
			\divide\count201 by 10
			\multiply\count201 by \count101
			\advance\count205 by \count201
		     \count201=\count200
			\divide\count201 by 100
			\multiply\count201 by \count102
			\advance\count205 by \count201
		     \edef\@result{\number\count205}
}
\def\compute@wfromh{
		\in@hundreds{\@p@sheight}{\@bbw}{\@bbh}
		\edef\@p@swidth{\@result}
}
\def\compute@hfromw{
	        \in@hundreds{\@p@swidth}{\@bbh}{\@bbw}
		\edef\@p@sheight{\@result}
}
\def\compute@handw{
		\if@height
			\if@width
			\else
				\compute@wfromh
			\fi
		\else
			\if@width
				\compute@hfromw
			\else
				\edef\@p@sheight{\@bbh}
				\edef\@p@swidth{\@bbw}
			\fi
		\fi
}
\def\compute@resv{
		\if@rheight \else \edef\@p@srheight{\@p@sheight} \fi
		\if@rwidth \else \edef\@p@srwidth{\@p@swidth} \fi
}
\def\compute@sizes{
	\compute@bb
	\if@scalefirst\if@angle
	\if@width
	   \in@hundreds{\@p@swidth}{\@bbw}{\ps@bbw}
	   \edef\@p@swidth{\@result}
	\fi
	\if@height
	   \in@hundreds{\@p@sheight}{\@bbh}{\ps@bbh}
	   \edef\@p@sheight{\@result}
	\fi
	\fi\fi
	\compute@handw
	\compute@resv}

\def\psfig#1{\vbox {
	\ps@init@parms
	\parse@ps@parms{#1}
	\compute@sizes
	\ifnum\@p@scost<\@psdraft{
		\special{ps::[begin] 	\@p@swidth \space \@p@sheight \space
				\@p@sbbllx \space \@p@sbblly \space
				\@p@sbburx \space \@p@sbbury \space
				startTexFig \space }
		\if@angle
			\special {ps:: \@p@sangle \space rotate \space}
		\fi
		\if@clip{
			\if@verbose{
				\ps@typeout{(clip)}
			}\fi
			\special{ps:: doclip \space }
		}\fi
		\if@prologfile
		    \special{ps: plotfile \@prologfileval \space } \fi
		\if@decmpr{
			\if@verbose{
				\ps@typeout{psfig: including \@p@sfile.Z \space}
			}\fi
			\special{ps: plotfile "`zcat \@p@sfile.Z" \space }
		}\else{
			\if@verbose{
				\ps@typeout{psfig: including \@p@sfile \space }
			}\fi
			\special{ps: plotfile \@p@sfile \space }
		}\fi
		\if@postlogfile
		    \special{ps: plotfile \@postlogfileval \space } \fi
		\special{ps::[end] endTexFig \space }
		\vbox to \@p@srheight sp{
			\hbox to \@p@srwidth sp{
				\hss
			}
		\vss
		}
	}\else{
		\if@draftbox{		
			\hbox{\frame{\vbox to \@p@srheight sp{
			\vss
			\hbox to \@p@srwidth sp{ \hss \@p@sfile \hss }
			\vss
			}}}
		}\else{
			\vbox to \@p@srheight sp{
			\vss
			\hbox to \@p@srwidth sp{\hss}
			\vss
			}
		}\fi

	}\fi
}}
\psfigRestoreAt
\let\@=\LaTeXAtSign

\title[Cosmochronology and the age of the galaxy]{Thresholds on star
formation and the 
chemical evolution of galactic discs: cosmochronology and 
the age of the galaxy}

\author[K. Chamcham and M. A. Hendry]
{K. Chamcham$^{1,2}$ and M. A. Hendry$^{1}$\\
 $^1$Astronomy Centre, University of Sussex, Falmer, Brighton, East Sussex 
BN1 9QH \\
$^2$University King Hassan II, Facult\'e des sciences 
                     Ain-Chok, B.P. 5366 Maarif, Casablanca, MOROCCO}

\date{Accepted ------. Received ------; in original form \today}

\begin{document}
\maketitle

\begin{abstract}
In this paper we analyse different chronometers based on
the models of chemical evolution developed in Chamcham, Pitts \& Tayler
(1993; hereafter CPT) and Chamcham \& Tayler (1994; hereafter CT). In those 
papers we discussed the ability of our models to reproduce the 
observed G-dwarf distribution in the solar neighbourhood,  
age-metallicity relation and radial chemical abundance gradients.
We now examine their response to the predictions of 
cosmochronology. We use the recent production ratios of the 
actinide pairs $^{235}$U/$^{238}$U 
and $^{232}$Th/$^{238}$U provided by Cowan, Thielemann \& Truran (1991) 
and the observed abundance ratios from Anders \& Grevesse (1989)
to determine the duration of nucleosynthesis in the solar neighbourhood, 
and thus to determine maximum likelihood estimates and
confidence intervals for the infall parameter, $\beta$, which controls
the growth rate of the disc in our models.
We compare our predictions for the age of the disc with the age of the 
galaxy estimated from models of white dwarf cooling and from the age of 
globular clusters. From our statistical analysis we find that these three
methods of age prediction appear to be consistent for a range of maximum
likelihood values of $\beta$ which is in good agreement with the values
considered in CPT and CT, which were found to give a good fit to the
observational data examined in those papers. We also briefly consider the
consistency of our results with the age of the universe predicted in
different cosmological models -- a topic which we will investigate more
fully in future work.
\end{abstract}


\section{Introduction}
\label{sec:introduction}
In a series of earlier papers, models for the chemical evolution of the galaxy
have been developed and studied by Chamcham, Pitts and Tayler based on the
adoption of a threshhold -- derived from the Toomre stability criterion --
for the onset of star formation (SF) in the galactic disc. In 
Chamcham, Pitts \& Tayler (1993; hereafter CPT) and Chamcham \& Tayler 
(1994; hereafter CT) the ability of these models to reproduce the observed 
G-dwarf distribution in the solar neighbourhood, age-metallicity relation and radial chemical abundance gradients was discussed. In this paper we examine 
the response of our models to the predictions of cosmochronology for the age 
of the galactic disc.

The age of the galaxy continues to be one of the most debated subjects in 
astronomy -- particularly in the light of a number of recent determinations of the Hubble constant (cf. Schmidt, Kirshner \& Eastman 1994; Freedman et al. 1994) which appear to indicate an age of the {\em universe\/} of less than 10 Gyr for the standard inflationary model of a flat universe with zero cosmological constant. These results are in
stark contradiction with astrophysical constraints on the age of the galaxy
from, e.g., the cooling of white dwarfs and the age of globular clusters
(cf. Renzini 1993). If both age determinations were correct then one would 
reach the absurd conclusion that the universe is younger than our galaxy, and 
the resolution of this paradox is one of the outstanding problems of
observational cosmology today.

In Table 1 we summarise the constraints on the age of the 
universe reported in recent literature, as provided by the four methods 
of age determination which we consider in this paper. Of course, of these
four methods, only by determining the values of cosmological parameters can 
one {\em directly\/} infer the age of the universe --
and even then only in the context of a given cosmological model. As indicated
in the first column of Table 1, the other three methods constrain the age
of the solar neighbourhood, galactic disc and galaxy respectively -- each of
which may then be regarded as a (progressively larger) lower bound for the
age of the universe. Thus, although the four methods are not
directly comparable, one can at least demand that their predictions should
be physically consistent with each other -- a useful approach which was 
adopted in, e.g., Tayler (1986) and which we also adopt in this paper. It 
would obviously be unreasonable for the age of the 
galaxy to be smaller than the age of the solar neighbourhood, or larger than 
the age of the universe, for example.

With respect to the second and fourth rows of Table 1, it would also
be naive to imagine
that all values in the cited range are equally likely. The age determined from
cosmological models extends to the limits of the quoted range only by a
somewhat delicate `balancing act' between the values of the relevant
cosmological parameters. In particular, if the Hubble constant is indeed
high -- say, $H_0 > 75$ -- then all but the extreme lower limit of the
quoted range is strongly excluded in the standard inflationary
model. Similarly, if
one believes the measured values and observational errors of the actinide
pairs' abundance ratios which we consider in this paper, then we shall see
that a careful statistical analysis can place tighter constraints on the
most likely age range determined from cosmochronology -- at least within 
the framework of a given chemical evolution model -- than suggested by
the full extent of the range quoted in Table 1.

In what follows we shall focus on the method of cosmochronology and 
compare its results with the predictions of the other methods.
Cosmochronology has
recently been gaining in credibility, especially with its extension to 
individual stars by Butcher (1987) --
although applications have usually been limited to the solar
system -- and the progress of theoretical and experimental nuclear physics
(Klapdor-Kleingrothaus 1991). This method is, however, still strongly 
dependent upon one's choice of chemical evolution model.

In CPT and CT we have shown that one of the weaknesses of our models is
the number of free parameters which we are using. The aim of this paper
is, therefore, to place some constraints on the most important of
these parameters through the analysis of cosmochronometers --
specifically, the abundance ratios of two actinide pairs -- by confronting 
their predicted age of the Galactic disc with other
astrophysical and cosmological predictions. The particular {\em strength\/}
of our models, on the other hand, is that the star formation (SF) law is 
deduced consistently with some physical input, 
instead of the simple Schmidt law (e.g. SF proportional to the gas density) 
commonly used by different authors in their cosmochronological models
(Schramm $\&$ Wasserburg 1970; Clayton 1988; Malaney \& Fowler 1987;
Pagel 1989).

Our present approach is analogous to that of Mathews and Schramm (1993)
in that both models consider a delay in SF instead of an instantaneous birth of
stars in the Galaxy. In our models this delay is related to the stability 
properties of the disc (cf. CPT), whereas in the model of Mathews $\&$ Schramm
it is introduced only schematically to sketch the merger process in the
formation of the Galaxy.
In our models this delay varies between 
1 -- 5 Gyr at the solar neighbourhood, depending on the growth time of the disc
(or equivalently the rate of infall of material onto the disc) and the 
relative roles of heating and cooling processes which regulate the star
formation rate (SFR). 

The structure of this paper is as follows. In section 2 we describe briefly 
our present chemical evolution model, summarising the relevant details from
earlier papers. In section 3 we discuss the results of applying our chemical
evolution model to predict the abundance ratios of the two chronometers as a
function of the age of the disc, and compare these model predictions with the 
observed abundance ratios. We then describe a method for obtaining maximum
likelihood estimates and confidence intervals for the infall parameter,
$\beta$, which controls the growth rate of the disc in our models, as a
function of the age of the disc. In section 4 we review in more detail
the constraints on the age of the galaxy provided from white dwarfs and 
globular clusters. In section 5 we then discuss the consistency of the
age predictions from our models with those from white dwarfs and globular
clusters, and on this basis consider what contraints are placed on the
parameter $\beta$, and on the age of the disc. We also briefly discuss
the consistency of our results with the age of the universe
predicted from the values of cosmological parameters -- a subject which
we will consider in more detail in a separate paper. Finally in section 6
we summarise our conclusions.

\section{Chemical evolution model}
\subsection{Threshold of star formation}
The determination of the age of the Galaxy by the method of cosmochronology 
provides at least a lower limit for the age of the universe. This method  is 
based on the
assumption that the earth and meteorites solidified some 4.5 Gyr ago and that
the isolation and condensation of the solar system from the ISM occurred at
$ T_{\odot}= 4.6\pm 0.1$ Gyr ago.
The present-day abundance ratios of the chronometric pairs
$^{235}$U/$^{238}$U and $^{232}$Th/$^{238}$U are determined from the analysis 
of meteorites (i.e. carbonaceous chondrites) and moon rock samples. 
Their values at the time of the formation of the solar system are 
extrapolated back in time and are respectively
$0.33\pm 0.03$ and $2.32\pm 0.23$ (Anders \& Grevesse 1989).

The duration of nucleosynthesis, $t_{\rm{nuc}}$, from the start of SF in the 
Galaxy until the formation of the solar system is determined by 
a suitable chemical evolution model. One can estimate from this a lower
limit for the age of the galactic disc to be
\begin{equation}
  T_{\tiny{D}} = T_{\odot} + t_{\rm{nuc}}
\label{Tgalaxy1}
\end{equation}
More correctly, equation \ref{Tgalaxy1} accounts for the age of the solar 
neighbourhood -- in other words the time since SF began at our location
in the Galaxy. We thus rename this estimated age $T_{\tiny{S}}$ instead of 
$T_{\tiny{D}}$.
\footnote{Note that $T_{\tiny{S}}$ is therefore comparable to the `raw' age
predicted by WD cooling.}

The variation in the models of cosmochronology used by different authors
stems from the different relations assumed between the SFR and the gas mass,
and their choice of model for the infall of material onto the disc
(Clayton 1988). To our knowledge all previous authors have assumed that SF
begins either instantaneously everywhere in the Galaxy or -- in the case of
Mathews \& Schramm (1993) -- after an arbitrary delay of a few Gyr. As 
shown in the threshold models of CPT, however, 
the delay, $\delta$, in the onset of SF in the solar neighbourhood is not an
arbitrary parameter but is determined self-consistently by the other
parameters of our chemical evolution model. We can therefore deduce a
model-dependent estimate for the age of the disc, $T_{\tiny{D}}$, given by
\begin{equation}
  T_{\tiny{D}}= T_{\odot} + t_{\rm{nuc}} + \delta
\label{Tgalaxy2}
\end{equation}

In the chemical evolution models developed in CPT and CT (respectively 
without and with radial flows), there is no reference to the process of 
formation of the halo and therefore no prediction of the time
delay between the formation of the halo and the disc. In other words, our
chemical evolution model does not allow a direct estimation of the age of the
halo, which we denote $T_{\tiny{G}}$, nor of course the age of the
universe, which we denote $T_{\tiny{U}}$, since our model places no constraints
on the `halo gestation period' for our galaxy -- i.e. the time elapsed from 
the Big Bang to the formation of the halo. 
As we already remarked in section 1, however,
for {\em{consistency}\/} we clearly require that the following 
chain of inequalities always holds:
\begin{equation}
 T_{\tiny{U}} > T_{\tiny{G}} \geq T_{\tiny{D}} \geq T_{\tiny{S}}
\label{ineq}
\end{equation}

In this paper we will derive constraints on the infall parameter,
$\beta$, of our chemical evolution model by comparing our model predictions 
for the age of the disc with other 
estimates of the age of the solar neighbourhood, the galactic halo and the 
universe, subject to the condition that the above inequalities should hold --
in other words that our chemical evolution model should be consistent with
other astrophysical and cosmological age determinations.

In order to determine the abundance ratios of the radioactive isotopes, we
follow the same procedure as in CPT, with the simple difference that
when SF starts, we solve the following equation for the evolution 
of the fraction of mass, $Z_{A}$, of a radioactive isotope $A$
(Tinsley 1980)
\begin{equation}
 {dZ_{A}\over d{t}} = {p_{A}s - {a} Z_{A} \over \Sigma_{g}} - 
             \lambda_{A}\;Z_{A}         
\label{metal}
\end{equation}
where $\lambda_{A}$ is the rate of disintegration of the isotope, $A$,
$p_{A}$ is its yield -- assumed to be constant -- $\Sigma_{g}$ is the gas
surface density, $s$ is the SFR and $a = a_0 f$ is the infall rate. We have
used the model infall function
\begin{equation}
f = 5 \beta \exp (-k r) [ \rm{sech} (\beta t) (1 + l \, {\rm{tanh}}
(\beta t) ) ] / (1 + {{2 l} \over {\pi}})
\label{infall}
\end{equation}
where the denominator is introduced as a normalisation factor 
(to ensure that the final mass of the galaxy is not dependent on either
$\beta$ or $l$) and $\beta$ is the infall parameter, as defined in CPT.
We have chosen the value of the parameter $a_0$ such that the final mass of
our model galaxy is in agreement with the observed range of values for 
our Galaxy.
In conjunction with equation \ref{metal}, we solve the equation for the gas
surface density,
\begin{equation}
{{d \Sigma_g} \over {dt}} = a - s
\label{gassurfdens}
\end{equation}

As in CPT, we require that SF starts only when the Toomre stability criterion
is marginally satisfied, i.e.:-
\begin{equation}
Q = {{\kappa c_g} \over {\pi G \Sigma_g}} \simeq 1
\label{toomre}
\end{equation}
where $\kappa$ is the epicyclic frequency and $c_g$ the gas velocity dispersion.
This threshhold on SF implies the delay, $\delta$, discussed above -- which
is partly dependent on the thermal properties of the disc and the
characteristics of the infall, amongst other physical properties (See CPT for
a more detailed discussion). In the present paper we concentrate on the
properties of the infall rate.

We write equations \ref{metal} and \ref{gassurfdens} in their 
non-dimensional form
\begin{equation}
 {dz_{A} \over d{\tau}} = { y - f \: z_{A} \over x } - \Lambda_{A} \: z_{A}
\label{metand}
\end{equation}
and
\begin{equation}
{dx \over d\tau} = f - y
\label{gsdndim}
\end{equation}
where we have used the notation defined in CPT; in particular here we 
have written $Z_{A}= p_{A}z_{A}$ . With this notation, the abundance ratio 
of an isotopic pair (A,B) with a production ratio $P_{A}/P_{B}$ is
\begin{equation}
{ N_{A} \over N_{B} } = { P_{A} \over P_{B} }{ z_{A} \over z_{B} }
\label{ratio}
\end{equation}

Here we have used the definition of the yield as the ratio of newly
synthesised elements returned instantaneously to the ISM, to the fraction 
of mass locked up in long-lived remnants: i.e.
$p_{A} =  A m_{u} P_{A} / (1 - R)$, where $R$ is the ejection rate and
$m_u$ is the atomic mass unit.
Furthermore we have defined $\Lambda_{A} = t_{0} \: \lambda_{A}$ as the 
dimensionless disintegration rate. 
Note that a fully consistent analysis should ideally consider a variable 
yield, especially at the early stages of the production of the radioactive 
elements (cf. Pagel 1993).

In order to calculate the initial (i.e. at the time when SF begins) 
enrichment of a given isotope, $A$, we assume in our models that there 
existed a `primordial' metallicity, $z_{i0}$, originating from earlier 
nucleosynthesis in the halo at an average time of
approximately $t_{e}$ Gyr before SF starts in the solar neighbourhood.
This generic assumption can find some support in the observed high 
($\approx 0.5$) [O/Fe] ratio for metal poor halo stars, 
indicating contamination from 
high mass stars formed during early SF in the galactic halo (Burkert 1993).
Under this assumption the initial metal enrichment would then be given by
$ z_{Ai}= z_{i0}\cdot \exp( - \lambda_{A} t_{e} ) $. To improve upon this
approximation we would need to further improve the models of CPT by a 
detailed study of the early moments of galaxy formation, in order to be able 
to follow the chemical evolution of the halo and the halo/disc interaction.
We will show in section 5, however, that the predicted abundance ratios at
late times have negligible dependence on the adopted values of $z_{i0}$
and $t_{e}$.

\subsection{Radial flows}

For the case where we have studied the effect of radial flows 
on predictions for the age of the disc, we have used the calculations 
developed previously in CT --
with the only change in the metallicity equation, in which we have added 
the term $ - \lambda_{A}\;Z_{A}\;\Sigma_g$ to the right hand side to account
for the decay of the radioactive element in the gas.
As was shown in CT, after a certain time of evolution of the disc 
when SF is well underway, incorporating radial flows in the model
will tend to produce a lower metallicity. Hence for a {\em given\/}
(i.e. the observed) metallicity one would expect the 
disc to appear {\em older\/} under the influence of radial 
flows than in their absence. This is because the fresh, metal-poor, gas 
flowing into the disc from the halo dilutes the heavy metals produced by
SF.

The equation of evolution of radioactive elements is thus,
\begin{equation}
\Sigma_{g} \left( {\partial Z_{A} \over \partial t} + 
             v_{r}{\partial Z_{A} \over \partial r} \right) = 
 p_{A} s - a Z_{A} - \lambda _{A} \Sigma_{g} Z_{A}
\label{rflow}
\end{equation}
Expressed in its non dimensional form this equation becomes
\begin{equation}
{\partial z_{A} \over \partial \tau} +
  V_{r}{\partial z_{A} \over \partial \bar{r}} = 
{ y - f z_{A} \over x } - \Lambda_{A} z_{A} 
\label{rflowbis}
\end{equation}
where $\bar{r} = r / r_0$, $V_{r} = v_r / v_{0r}$, $v_{0r} = r_0 / t_0$,
with $r_0 = 1$ kpc and $t_0 = 1$ Gyr.

At this stage we need to treat carefully the solution of the above equation.
When we include radial flows all regions of the
disc are intercorrelated as star formation propagates outwards
from the centre of the disc (see figure 1, where we plot the time, 
$t_{\star}$, at which SF begins at radius, $R_{\star}$) and ideally we should 
account for this when modelling the initial metal enrichment of the solar 
neighbourhood. A fully consistent treatment to this problem is far from
trivial, however, and -- as we indicated previously -- would require a 
detailed model for the dynamical evolution of the halo, both prior to the
onset of SF in the disc and throughout the subsequent evolution of the disc.
We can, nevertheless, place useful limits on the impact of including radial
flows on our age predictions for the disc by considering the case where the 
initial metal enrichment is identically equal to zero. Since the effect of
radial flows will always be to dilute the primordial metallicity, $z_{i0}$,
setting the initial enrichment equal to zero when SF begins at each radius in
the disc is simply the limiting case of this dilution process.

\section{Results}

We now describe in detail the numerical results of solving the equations of our
chemical evolution models outlined in section 3. Our solution involves the
adoption of numerical values for the following three parameters: $K_1$, $K_2$ 
and $K_3$, which denote the fraction of energy put into the ISM by infall, SF
and cloud-cloud collisions respectively. We have used the following 
numerical values (arguments for the choice of which were given in CPT):
\begin{equation}
 K_{1} = 2 \, , \;\;  K_{2} = 25 \, , \;\; K_{3} = 10
\end{equation}

The values of the (dimensionless) disintegration constants which we adopt are 
$\Lambda_{A} \equiv $ 0.0495 for $^{232}$Th, 0.985 for $^{235}$U and 
0.1551 for $^{238}$U (cf. equations \ref{metand} and \ref{rflowbis}). Varying
the remaining free parameters of our models, we found that the predicted
abundance ratios are sensitive only to the infall parameter, $\beta$. We thus
chose to address quantitatively the sensitivity of our age predictions to
$\beta$ and assigned to all other model parameters the same numerical values 
as in `Model 1' of CPT (See Table 1 of CPT for further details).

Figure 2 illustrates the dependence of SF on the infall parameter, 
plotting the SFR as a function of time
for $\beta =$ 0.1, 0.4, 0.7 and 1.0, and it is clear that the value of $\beta$
determines not only the delay, $\delta$, before the onset of SF, but also
the duration and amount of SF in the solar neighbourhood, at a given time.
For $\beta = 0.1$ the onset of SF is delayed by $\delta \sim 6$ Gyr, and SF then
proceeds as a slowly varying function of time. For $\beta = 1.0$, on the other 
hand, the delay, $\delta$, is less than 1 Gyr, and is followed by a rapid burst
of SF which has decayed to a few percent of its peak rate after only $\sim 7$
Gyr.
  
In order to address quantitatively this sensitivity to $\beta$, we consider 
below a statistical
analysis designed to obtain maximum likelihood estimates of $\beta$, and thus
constrain the law of SF, which `best fits' our chemical evolution model to the 
observational age constraints in a manner consistent with the inequalities of
equation \ref{ineq}.

Among other tests to which one can subject our models are their ability to
solve the G-dwarf problem, to fit the age-metallicity relation and to predict 
the observed gas fraction in the solar neighbourhood. The relative success of 
the models in addressing these problems has already been discussed in CPT.
In figure 3 we illustrate the explicit dependence of the gas fraction, $\mu$,
on the adopted value of $\beta$ -- associated with the SFR as shown in 
figure 2. We plot the model-predicted values of $\mu$ as a 
function of the present age of the galactic disc, and compare them to 
the observed value of $\mu_{\rm{obs}} = 0.28 \pm 0.09$, taken from Gilmore,
Kuijken \& Wyse (1989), and denoted by the horizontal lines on figure 3.
We also illustrate, shown on the upper ordinate of figure 3, the 
relationship between the model-predicted values of $\mu$ and the age of the
{\em universe\/} -- a lower limit for which is estimated by adding 1.5 Gyr 
to account for the time interval between the Big Bang and the formation of 
the galactic disc.

We can see from the figure that the predicted value of $\mu$ matches the 
observed value at a wide range of different disc ages, depending on the 
value of $\beta$, ranging from $\sim 20$ Gyr for $\beta = 0.1$ to 
$\sim 3$ Gyr for $\beta = 1.0$. N ote that for 
$\beta > 0.4$ the best fit of the model predictions to the observations is 
found for a low disc age of $t < 6.5$ Gyr -- in good agreement with the age
of the universe predicted for the standard inflation model with a fairly 
high ($H_{0} \sim 70$) Hubble constant, even if our adopted disc `gestation 
period' of 1.5 Gyr were something of an underestimate. The shape of the curves 
and the width of the observational error band on $\mu_{\rm{obs}}$, however, 
mean that considerably higher disc ages are not strongly excluded -- 
even for $\beta = 1$. 

We have computed the abundance ratios of the chronometric pairs 
$^{232}$Th/$^{238}$U and $^{235}$U/$^{238}$U with the respective production
ratios $1.6 \pm 0.1$ and $1.24 \pm 0.1$ from Cowan, Thielemann \& Truran
(1991). 
Our computations show that the quoted uncertainty of $\pm 0.1$
on the production ratios introduce a spread of about $\pm 1$ Gyr
on the predicted age, in the sense that the higher the production ratio the
younger the disc and vice-versa.

Figures 4a and 4b show the time evolution of the model predicted values
of the abundance ratios of our two chronometers -- again associated with
the same
SFR of figure 2. As in figure 3, the observed abundance ratio and its 
$1 \sigma$ error band are denoted by the horizontal lines on each figure.
These graphs have been obtained adopting a delay of
$t_e = 3$ Gyr between the period of nucleosynthesis in the halo -- which we
assume resulted in a `primordial' metallicity of $z_{i0} = 0.25$; see section 
2 -- and the onset of SF in the solar neighbourhood. Reasons for the choice of 
$z_{i0} = 0.25$ are discussed in CPT; we show in section 5, however, that the
adopted values of both $z_{i0}$ and $t_e$ have negligible impact on our
results.

Note that, in a similar manner to figure 3, we estimate the time evolution of 
the production ratios in our model as a function of the age of the 
universe (as shown in the upper ordinates of figures 4 and 5) by 
adding 6 Gyr to the age of the disc: a disc gestation period of 1.5 Gyr 
and a further 4.5 Gyr representing the interval from the formation of the solar 
system to the present day -- during which time the abundance ratios have 
been `frozen'. As in the case of figure 3, this second age scale is designed 
to be largely illustrative -- simply in order to allow an approximate lower
bound for the age of the universe to be estimated from our models.

A quantitative examination of figures 4a and 4b reveals several 
interesting features. Firstly, note that the 
initial value of the $^{232}$Th/$^{238}$U ratio predicted in our model lies 
within the $1 \sigma$ error band of the observed value of this ratio, and only
deviates significantly from the observed ratio several Gyr after the onset of
SF (the exact time depending on the value of $\beta$) before reaching a 
minimum value
and then increasing monotonically at later times. The $^{232}$Th/$^{238}$U 
abundance ratio, on the other hand, is predicted in the model to have an initial
value several standard deviations below the observed ratio, to increase rapidly
after the onset of SF, to reach a maximum several Gyr later and then to decay
monotonically at later times. 

A naive interpretation of these figures could then
lead to the conclusion that our model predictions give a good fit to the 
observed abundance ratios for very young ages. For example, in the case of
$\beta = 0.4$ we see from figure 4b that the ratio $^{232}$Th/$^{238}$U 
rises sharply to intercept the observed abundance ratio when the age of the
disc is only 2 Gyr.
At the corresponding age on figure 4a we see that the model
predicted ratio of $^{232}$Th/$^{238}$U has decreased only very slightly from
its initial value, and still lies comfortably within the $1 \sigma$ error band.
Thus, when $\beta = 0.4$, it is apparent that the observed and predicted 
abundance ratios are in good agreement for a disc which is only 2 Gyr old.
The obvious flaw in this interpretation, however, is the fact that SF has 
begun only 1 Gyr before the formation of the solar system with these model
parameters. In other words this apparently good fit to the observed abundance
ratios does not allow a nearly sufficient duration for nucleosynthesis. We 
know, from the ages of WD in the solar neighbourhood, that significant amounts 
of SF had occured in
our region of the galactic disc around 9 Gyr ago\footnote{See also section 4
for a more detailed discussion}. Allowing 4.5 Gyr for the age of the solar 
system, this means that we should allow a minimum of $\sim 4.5$ Gyr to account 
for the duration of nucleosynthesis - i.e. the interval, $t_{\rm{nuc}}$, 
between the onset of SF in the solar neighbourhood and the formation of the 
solar system. In particular, therefore, any assessment of the `goodness of fit' 
between our model predictions and the observed abundance 
ratios must ensure that a sufficiently long period of nucleosynthesis is 
allowed. Similarly, a statistical analysis designed to estimate `best fit'
values of the infall parameter, $\beta$, should exclude all cases in which the
duration of nucleosynthesis is unreasonably short. We now describe such a
statistical analysis.

We begin by introducing some notation. Let ${\rm U_{obs}}$ and ${\rm Th_{obs}}$
denote the observed abundance ratios of $^{235}$U / $^{238}$U and 
$^{232}$Th / $^{238}$U 
respectively, and let ${\rm U_{true}}$ and ${\rm Th_{true}}$ denote the true 
values of these ratios. We define ${\rm \varepsilon_{\tiny{U}}}$ and 
${\rm \varepsilon_{\tiny{Th}}}$, the observational errors on the measured 
values of the production ratios, as:

\[ {\rm \varepsilon_{U} = U_{obs} - U_{true} } \] 

\[ {\rm \varepsilon_{Th} = Th_{obs} - Th_{true} } \] 

\noindent
Finally we define ${\rm U_{model}}(t,\beta)$ and ${\rm U_{th}}(t,\beta)$ as the
present day values of $^{235}$U / $^{238}$U and $^{232}$Th / $^{238}$U predicted
in our model as a function of the age of the universe, $t$, and the value of the
infall parameter, $\beta$. We wish to construct a joint likelihood distribution,
${\cal L}(t,\beta)$, for $t$ and $\beta$ under the null hypothesis that our
model is correct. In other words, ${\cal L}(t,\beta)$ measures the probability 
that one would obtain the observed values of the abundance ratios, given that
their true values are equal to the values predicted in our model -- 
for given $t$ and $\beta$. Thus, under the null hypothesis, we have:

\[ {\rm \varepsilon_{\tiny{U}} = U_{obs} - U_{model} }(t,\beta) \] 

\[ {\rm \varepsilon_{\tiny{Th}} = Th_{obs} - Th_{model} }(t,\beta) \] 

We assume that ${\rm \varepsilon_{\tiny{U}}}$ and 
${\rm \varepsilon_{\tiny{Th}}}$ are normally
distributed with zero mean and dispersion equal to the standard error of the
respective abundance ratio, as reported in the literature. 
We also assume that ${\rm \varepsilon_{\tiny{U}}}$
and ${\rm \varepsilon_{\tiny{Th}}}$ are uncorrelated - which is certainly not 
the case for the values of ${\rm U_{true}}$ and ${\rm Th_{true}}$, but is 
likely to be a reasonable approximation for their observational errors.

The joint likelihood function, ${\cal L}(t,\beta)$, is then given by:
\begin{eqnarray}
&  &
 {\cal L}(t,\beta) dt d\beta = A \exp \left[ - {1 \over 2}
\left( {\rm U_{obs} - U_{model}}(t,\beta) \over \sigma_{\tiny{U}} \right)^{2} - 
\right.                  \nonumber \\
&  &
\left. {1 \over 2} 
\left( {\rm Th_{obs} - Th_{model}}(t,\beta) \over \sigma_{\tiny{Th}} \right)^{2}
 \right] S(t,\beta) dt d\beta
\label{likeli}
\end{eqnarray}
Here ${\rm \sigma_{\tiny{U}}}$ and ${\rm \sigma_{\tiny{Th}}}$ denote the 
dispersion of ${\rm \varepsilon_{\tiny{U}}}$ and 
${\rm \varepsilon_{\tiny{Th}}}$ respectively, $A$ is a
normalisation constant and $S(t,\beta)$ is a selection function which excludes
values of $t$ and $\beta$ which are physically unreasonable due to, e.g., too
short a duration for nucleosynthesis.   

It now follows from equation (\ref{likeli}) that the likelihood distribution of
$t$ conditional upon $\beta$ is given by
\begin{equation}
 {\cal L}(t | \beta) dt = { {\cal L}(t,\beta) dt \over 
\int {\cal L}(t,\beta) dt }
\label{likelinormt}
\end{equation}

In figure 5 we plot this distribution for the illustrative values of 
$\beta = 0.1, 0.4, 0.7, 1.0$, as before. We assume a {\em minimum\/} duration, 
$t_{\rm{nuc}}$, of nucleosynthesis of 4.5 Gyr. Of course the minimum 
acceptable age 
for the disc at the formation of the solar system is in practice higher than 
4.5 Gyr, since our model also predicts a $\beta$ dependent time delay, $\delta$,
before the onset of SF in the solar neighbourhood, as we have already seen in 
figure 2, and this limiting age causes the slight skewness in the likelihood 
distribution for 
$\beta = 0.7$ and 1.0 in figure 5. We see that for $\beta = 0.1$ we require an
age for the disc at solar system formation of between $\sim 13$ Gyr and 
$\sim 24$ Gyr
to match the observed abundance ratios with a non-negligible likelihood,
while for $\beta = 1.0$ the likelihood is non-negligible over the
much narrower range of $\sim  5.3 - 7$ Gyr. Note that, while the likelihood 
distributions for
$\beta = 0.1$ and $0.4$ are effectively disjoint, there is a significant
overlap between the likelihood distribution for $\beta = 0.4$ and 0.7 and
for $\beta = 0.7$ and 1.0. This
reflects the fact that the likelihood distributions become increasingly
`pushed up' against the lower limit of $t_{\rm{nuc}} \geq 4.5$ Gyr as 
$\beta$ increases.
Another consequence of this is the fact that as $t$ decreases, the value of 
$\beta$ is less well constrained by the observed abundance ratios, since there
is a wider range of values of $\beta$ for which  ${\cal L}(t | \beta)$ is
comparable in magnitude.

We can illustrate more precisely the constraints on $\beta$ as a function 
of $t$
by considering the conditional likelihood, ${\cal L}(\beta | t)$, given by
\begin{equation}
 {\cal L}(\beta | t) d\beta = { {\cal L}(t,\beta) d\beta \over 
\int {\cal L}(t,\beta) d\beta }
\label{likelinormb}
\end{equation}
 
Figure 6 shows likelihood curves for ${\cal L}(\beta | t_{\rm{disc}})$ 
for several values of $t_{\rm{disc}}$, the age of the disc at the formation
of the solar system. It is clear that the likelihood distribution does
indeed become slightly narrower as the age of the disc increases. Using 
equation \ref{likeli} it is straightforward to construct a curve of the maximum 
likelihood estimate, $\hat{\beta}_{ML}(t_{\rm{disc}})$, of $\beta$
as a function of $t_{\rm{disc}}$. We plot this curve in figure 8. 
We can see that if $t_{\rm{disc}}$ lies in the range e.g. 8 - 12 Gyr, this
corresponds to a maximum likelihood estimate of $\beta$ lying in the 
approximate range 0.25 - 0.45. 

Of course the form of the likelihood curves plotted in figures 5 -- 7
depends in part upon our adopted value for the minimum duration of
nucleosynthesis -- so far taken to be
4.5 Gyr. Increasing $t_{\rm{nuc}}$ to, e.g., 6 Gyr causes a further increase 
in the skewness of the likelihood distributions,
${\cal L}(t | \beta)$ for large values of $\beta$. The likelihood distributions
conditional upon age, ${\cal L}(\beta | t)$, are largely unaffected by the
increase in $t_{\rm{nuc}}$, however, although of course there is now no 
likelihood curve for a disc age of 6 Gyr as this value is excluded by the 
higher value of $t_{\rm{nuc}}$ because one must also allow for the $\beta$
dependent time delay before the onset of SF in the solar neighbourhood. 
Figure 8 shows the maximum likelihood estimate of $\beta$ as a function of 
$t_{\rm{disc}}$, but now with $t_{\rm{nuc}}$ constrained to be at least 6 Gyr.
We see that there now exists no acceptable solution for 
$\hat{\beta}_{ML}(t)$ for a disc younger than $\sim 7$ Gyr. In other words, 
it is not possible for our model to build the disc quickly enough and still 
allow at least 6 Gyr between the onset of SF and the formation of the Solar 
System. Adopting our estimate of 1.5 for the disc gestation period, this
translates into a lower limit on the age of the universe, $T_{\tiny{U}}$, of
13 Gyr. If $T_{\tiny{U}}$ lies in the interval $\sim 13 - 13.5$ Gyr, the 
effect of increasing the lower limit on $t_{\rm{nuc}}$ to 6 Gyr is to force a 
sharp increase in the value of $\hat{\beta}_{ML}(t)$, in order
that SF begins {\em early\/} enough to give a sufficiently long duration for
nucleosynthesis. Above an age of $\sim 13.5$ Gyr, however, 
 $\hat{\beta}_{ML}(t)$ is identical to the value obtained with the smaller
lower limit of $t_{\rm{nuc}} \geq 4.5$ Gyr, as shown in figure 7 -- a fact 
which is easily verified analytically from differentiation of equation 
\ref{likelinormb}.

Following the method described in Hendry $\&$ Simmons (1990)
we can construct confidence curves for the infall parameter, $\beta$, and the
age of the disc, $t_{\rm{disc}}$, when the solar system formed. From 
these curves we can derive confidence intervals for $\beta$ given 
$t_{\rm{disc}}$, or $t_{\rm{disc}}$ given $\beta$. Examples of such 
curves are shown in figures 9a and 9b, where the confidence intervals are 
derived at 67\% and 90\%, and with an assumed lower limit of 
$t_{\rm{nuc}} \geq 4.5$.
By drawing a vertical line at some fixed value of $t_{\rm{disc}}$ the points
at which the line intercepts the two curves define the limits of the
appropriate confidence interval for $\beta$. 

We can see from figures 9a and 9b that as 
$t_{\rm{disc}}$ increases, the confidence intervals for $\beta$ become 
somewhat narrower in width and the limits of the intervals are monotonically
decreasing -- i.e. the estimated value of $\beta$ is both smaller
and more tightly constrained for larger values of $t_{\rm{disc}}$.
We can see, moreover, that for $t_{\rm{disc}} < 8$ Gyr, the confidence 
intervals are pushed sharply to high values of $\beta$. This
effect is also dependent on the adopted lower limit for $t_{\rm{nuc}}$, and
can be seen to be somewhat more pronounced for the confidence curves
with $t_{\rm{nuc}} \geq 6$ Gyr, as shown in figures 10a and 10b. 
Clearly a high value of $\beta$ is required if SF is to begin in the solar
neighbourhood quickly enough to allow a disc age of less than 8 Gyr when the
solar system is formed and at the same time allow a sufficient duration for
nucleosynthesis -- particularly with $t_{\rm{nuc}} \geq 6$ Gyr. As was the
case with figures 7 and 8, on the other hand, the adopted lower 
limit on $t_{\rm{nuc}}$ has no effect on the shape of the confidence curves 
for $t_{\rm{disc}} > 10$ Gyr.

In order to constrain $\beta$ more precisely from this analysis --
and to determine if the age predictions of our model can be made
consistent with the
inequalities of equation \ref{ineq} -- we now consider what are the
constraints upon the age of the disc from estimates of white dwarf cooling
and globular clusters.

\section{Constraints on the age of the galactic disc from other methods }
\subsection{White dwarfs}

On the basis of the theory of cooling of white dwarfs (WD), developed by Mestel 
(1952), Schatzman (1958) and Mestel $\&$ Ruderman (1967), which gives 
a relation
between their age and luminosity, one can estimate the age of the galactic disc
by measuring the luminosity of the coolest  white dwarfs (Schmidt, 1959). This
method requires the adoption of an equation of state for white dwarfs,
knowledge of their 
chemical composition (i.e. the relative abundance of C, O, He and H)
and the fluid -- crystal phase transition, since the cores of white dwarfs are
supposed to crystallize at low 
luminosity.  We do no more than summarise some salient points from
recent applications of this method. For a comprehensive review see e.g.
D'antona \& Mazzitelli (1990).

Hotter WD cool more rapidly, therefore the space-density of WD is expected to 
increase monotonically with decreasing WD luminosity. On the other hand, 
because of the finite age of the galactic disc there should be a paucity 
of WD at low luminosity. This is confirmed by observations which show an 
abrupt fall off in 
the number of WD stars below $\log(L/L_{\odot}) = -4.5$. Therefore, because the
WD are considered to be the oldest stars in the disc, the age of the galactic 
disc can be inferred by computing the time required for them to cool to this
luminosity, with a correction to account for their main sequence (MS) lifetime. 

Winget et al. (1987) have fitted the observational data of WD luminosities
with a model assuming a
pure carbon core and a constant birthrate of WD over the age of the Galaxy. 
They deduced an age of the galactic disc $T_{\tiny{WD}}= 9.3 \pm 2$ Gyr.
They then adopted an estimate for the 
age of the universe of $T_{\tiny{U}}= 10.3 \pm 2.2$ Gyr, which they claim
accounts for the ``time between the Big Bang and the first appearance of
stars in the Galactic disk". It seems to us that their addition of
only 1 Gyr to arrive at an estimate of $T_{\tiny{U}}$ is at best somewhat
arbitrary and at worst a considerable underestmate, given that their 
$T_{\tiny{WD}}$ is inferred only from WD in the solar neighbourhood and 
hence only takes account of the history of SF at our location in the disc. 
Their value of $T_{\tiny{WD}}= 9.3 \pm 2$ Gyr {\em does\/}, however, provide 
a reasonable estimate of the age which we denote by $T_{\tiny{S}}$ in 
equation \ref{ineq} -- the age of the solar neighbourhood.

Iben \& Laughlin (1989) considered in their model more ingredients,
such as the IMF, time-variable SFR and the 
explicit influence of the lifetimes of MS progenitors. They found that the shape
of the luminosity function is independent of any variation in the SFR, 
supporting thus the assumption of Winget et al. They deduced an age for the 
galactic disc of $T_{\tiny{WD}} = 9-10$ Gyr, consistent with the Winget et al.
result.

As with the Winget et al. study, we regard this result as a reliable estimate
of $T_{\tiny{S}}$. Both determinations should be regarded as no more than a 
lower limit for the present age of the galactic disc, $T_{\tiny{D}}$, and
the age of the galaxy, $T_{\tiny{G}}$, for the following reasons:
\begin{enumerate}
\item{disc stars in the solar neighbourhood form certainly after the halo 
and bulge stars with a time delay of a few Gyr}
\item{not all the stars born in the disc stay there,
especially the first generation of stars which inherit high random
velocities from the gas out of which they formed. These stars therefore undergo 
more scattering, contributing thus to the WD paucity observed}
\item{any peculiar phenomenon liberating energy (nuclear, gravitational etc.) 
will tend to prolong the cooling time, and therefore increase the age of the 
galactic disc obtained by this method.}
\item{ there is no reason why the evolution of WD should be independent of
the chemical evolution of the disc, a fact pointed out in Pitts \&
Tayler (1992). In fact Pitts $\&$ Tayler 
showed, from their chemical evolution model, that the study of WD
may be an underestimate of the true age: they deduced an age $\sim 1$ Gyr
higher than that predicted by Winget et al.}
\end{enumerate}

\subsection{Globular Clusters}

Globular clusters (GC) are known to be the oldest observed stellar systems in 
our Galaxy -- and to some extent in the Magellanic Clouds. All studies 
have generally converged upon an age in the range 
$T_{\tiny{GC}} =  13 - 15 \pm 3$ Gyr (VandenBerg 1990; Renzini 1993), 
although the age probability distribution 
is highly skewed and renders ages below 12 Gyr very unlikely (Tayler 1986,
Renzini 1993).

The age of GC can be determined by comparing their observed H-R diagrams with 
theoretical H-R diagrams calculated as a function of time. Assuming that all 
the observed stars in the cluster have been formed on a time-scale which is 
short compared to the cluster age, and that all stars are formed from matter 
of the same composition, a direct measure of the age of the cluster can be 
obtained if the luminosity of the MS stars of given composition is known at 
the turnoff point (TO) -- the position on the
HR diagram which nearly coincides with the exhaustion of H at the centre of
evolving stars -- by using the time-luminosity relation for the MS-TO 
(VandenBerg 1988).
The two methods commonly used to obtain the TO luminosity involve one of the
following:
\begin{enumerate}
\item{the direct MS fitting of observed colour-magnitude diagrams to
theoretical models} 
\item{the calibration of the magnitude difference between H-B stars and the MS 
turnoff, at the colour of the turnoff.}
\end{enumerate}

The accuracy of either method depends upon how realistic is the 
theoretical H-R diagram used (Chiosi, Bertelli \& Bressan 1992), 
as well as the extent of the uncertainties relating to the observed chemical 
compositions, reddening and distance moduli (Renzini 1993).
It is these systematic effects which chiefly contribute to the quoted
systematic uncertainty of $\pm 3$ Gyr in Renzini (1993).

The GC 47 Tucanae has been the most studied and thus has the most reliably
determined age (Hesser, Harris \& VandenBerg 1987). 
Assuming an initial helium content of $Y = 0.24$ for the
GC stars, its age was found to be $\;\;T_{\rm{47Tuc}}\simeq 13.5 \pm 2\;$ Gyr. 
Note 
that the determination of the age of the GC is very sensitive to the 
uncertainty
in the He abundances, whose minimum values should be bound by the primordial
Big Bang  abundance (Tayler 1986). 
VandenBerg argues that an uncertainty as small as $\Delta Y = \pm 0.02$
has considerable ramifications for GC ages. There is no clear age-metallicity 
relation for GC, however, even if the choice of the metallicity seems 
to have a great 
influence on the determination of the age -- as shown in the case of the 
GC M92. Adopting a composition of $Y= 0.24$, $[Fe/H]= -2.03$ and 
$[O/Fe]= 0.70$ for M92 yielded an estimated age of 
$T_{92} \sim 14$ Gyr.
Adopting $[Fe/H]= 0$, with the other ratios unchanged,  gave
$T_{92} \sim 17$ Gyr. One is forced to conclude that the absolute
value of the age depends critically on the assumed chemical composition of the
member stars. Hence any improvement in GC age determination relies, in part, 
on improved determinations of their chemical composition.
Another important source 
of uncertainty is the stellar evolutionary model adopted. 
As an example, Vandenberg 
has shown that the inclusion of Helium diffusion -- which has been shown
to be an important process in Pop II stars -- in the canonical
models of stellar evolution can reduce the age of the GC by about $15\%$ --
pushing back towards the age predicted by WD.

Chaboyer (1995) has also attempted to study the influence on the GC age 
determination 
of the uncertainties in the input physics used in stellar evolution models, 
and concluded that the spread in `acceptable' GC age determinations -- given
the systematic errors -- could be at most 11 - 21 Gyr. The two major sources of 
error identified in this study originate from the difficulty of modelling
convection in stellar evolution models (e.g. changing the mixing length 
from 1.5 to 3 added an uncertainty in the age of up to 10\%),
and from the uncertainties in distance 
determinations to GC. Currently, distance moduli cannot be determined to an
accuracy of better than $\pm 0.2$ magnitudes, which contributes an
uncertainty of up to 22\%
to the inferred GC age (Renzini 1993). Improvements in the calibration of
each of the distance 
indicators commonly used -- RR Lyrae variables, Subdwarfs and WD --
are expected to come from the HST and the Hipparcos satellite in the near
future, however.

In summary, if we assume that GC are of a comparable age to that of the galaxy
itself, then GC ages essentially provide a direct measure of $T_{\tiny{G}}$
in equation \ref{ineq}. Following Renzini (1993) and Chaboyer (1995), we adopt
the estimate $T_{\tiny{G}} = 13 - 15 \pm 3$ Gyr for the purposes of
comparison with our chemical evolution models -- with the important
proviso that ages of {\em less\/} than 12 Gyr are highly unlikely.

\section{Discussion}

We now turn to the main question of this paper: how do
the predictions of our chemical evolution models
fit into the picture of other predictions for the age of the disc, and what
constraints do they allow us to place on the value of $\beta$?

We have seen in figures 7 and 8 that the maximum likelihood
value of $\beta$ decreases monotonically as the age of the galactic disc
increases. Suppose that the age of the disc at the formation of the Solar 
System is taken to be 9 Gyr -- which would imply an estimate for 
$T_{\tiny{U}}$ of $\sim 15$ Gyr, assuming a reasonably short gestation period 
of 1.5 Gyr for the disc. This value would also be in agreement with the age
predictions of globular clusters discussed above. In this case the 
maximum likelihood estimator of $\beta$, $\hat{\beta}_{\rm{ML}}$ = 0.39 
(cf. figure 8), and a 67\% confidence interval for $\beta$ is found from
figure 9a to be
$0.34 \leq \beta \leq 0.46$ (assuming $t_{\rm{nuc}} \geq 4.5$ Gyr; 
the confidence
interval is very slightly wider if we require $t_{\rm{nuc}} \geq 6.0$ Gyr).
We can see from figure 2 that, for this maximum likelihood value of $\beta$,
SF began in the solar neighbourhood about 2 Gyr after the formation of the
disc -- i.e. about 11.5 Gyr ago, which is also perfectly consistent with the
constraints on $T_{\tiny{S}}$ from WD cooling. We can see, moreover, from 
figure 3 that $\beta = 0.4$ gives a reasonable fit to the observed gas
fraction, for $T_{\tiny{D}} = 13.5$ Gyr. Recall also from CPT that this 
value of $\beta$ was successfully able to fit the observed age-metallicity 
relation and to solve the G-dwarf problem. Thus, a value of $\beta \simeq 0.4$ 
would seem to be favoured by our models -- predicting as it does a disc age 
which is consistent with the acceptable ranges for the other disc age 
predictions which we have considered.

The biggest difficulty in achieving consistency with the inequalities of
equation \ref{ineq} when $\beta = 0.4$ would, therefore, seem to be
finding a viable 
cosmological model which allows $T_{\tiny{U}} \geq 15$ Gyr. This is not an 
easy task, however, if $ \rm H_0$ lies in the range of 60 -- 70 km s$^{-1}$ 
Mpc$^{-1}$, as a number of recent reliable determinations have reported (c.f. 
Freedman et al. 1994; Schmidt, Kirschner \& Eastman 1994; Riess, Press \& 
Kirschner 1995). Such a high value of $\rm H_0$ would certainly rule out the
standard inflationary universe, with $\Omega_0 = 1$ and $\Lambda_0 = 0$.
Adopting the lower bound for the matter density, $\Omega_{\rm{m}} \geq 0.35$,
deduced in Liddle et al. (1995), we find that we {\em can\/} have
$T_{\tiny{U}} \sim 15$ Gyr provided $\Lambda_0 \neq 0$, an issue previously
examined in Tayler (1986) in which a similar conclusion was reached. Thus,
it would appear that constraints on cosmological parameters still allow the
consistency of {\em all\/} of the inequalities in equation \ref{ineq} for
a value of $\beta \simeq 0.4$, although certainly the range of compatible 
values of $\rm H_0$, $\Omega_0$ and $\Lambda_0$ is severely limited. In
Hendry \& Chamcham (1995, in preparation) we investigate in detail the
current observational status of these cosmological parameters, and
address specifically the issue of how readily one can match the consistent 
age predictions for the galactic disc which we have discussed in this paper
with the age of the universe predicted in cosmological models.

We now consider the sensitivity of our results to the adopted value of the
primordial metallicity, $z_{i0}$. Figure 11 shows the 67\% confidence curves
for $\beta$ and the age of the disc, assuming $t_{\rm{nuc}} \geq 4.5$ Gyr
and now with $z_{i0} = 0$. We can see immediately, on comparison with
figure 9a, that for {\em all\/} values of $\beta$ the change in the
confidence curves is negliglible. Specifically, the confidence limits for
the age of the disc at a given value of $\beta$ are increased by
$\sim 0.1$ Gyr (or equivalently the confidence limits for $\beta$ at a given
disc age are increased by $\sim 0.01$) when $z_{i0}$ is decreased from
0.25 to the limiting value of zero. Although the magnitude of the
difference in the confidence limits between figures 9a and 11 is 
clearly negligible, it is worth noting that the {\em sign\/} of the 
difference is certainly in accordance with our physical understanding:
when the primordial metallicity is reduced, one would expect to need a
slightly longer time to build the observed heavy element abundance ratios,
for a given value of $\beta$.

Note that figure 11 also demonstrates that
our results are completely insensitive to the adopted value of $t_e$:
setting $z_{i0} = 0$ is numerically equivalent to letting $t_{e}$ tend to
infinity. Although very large values of $t_e$ would, of course, not make 
physical sense 
-- as in any case $t_e$ is constrained to be shorter than the time interval 
between the Big Bang and the onset of SF in the disc -- it is, nevertheless, 
clear that the chosen value of $t_e$ is essentially unimportant to our 
analysis.

Finally we consider the effect on our results of including radial flows
in our chemical evolution models. As we explained in section 2, a completely
self-consistent treatment of this problem would require a detailed model for
the dynamical evolution of the halo. Instead, in this paper, we again
consider the limiting case where $z_{i0} = 0$ everywhere in the disc at
the onset of SF. Thus we explicitly include the effects of radial flows in the
solution of equation \ref{rflowbis} only {\em after\/} SF begins, which is 
essentially equivalent to assuming that the
inflow of fresh, metal-poor, gas from the halo prior to the onset of SF
{\em completely\/} dilutes the metal content of the disc during that time. 
Figure 12 shows the 67\% confidence curves
for $\beta$ and the age of the disc in this case, and assuming 
$t_{\rm{nuc}} \geq 4.5$ Gyr. We can see from this figure that, for a disc
age of 9 GYr at the formation of the Solar System, the 67\% confidence
interval for $\beta$ is now $0.40 \leq \beta \leq 0.52$ -- i.e. the
upper and lower confidence limits for $\beta$ are increased by 0.06.

This small positive shift in the confidence interval for $\beta$ 
is again in accordance with our physical expectations: in the radial flow
model, the continual dilution of the disc with metal-poor gas from the halo 
would imply that we need a slightly higher value of $\beta$ in order to
build the observed heavy element abundances in a given time.
The shift is clearly not a large effect, however, particularly 
since a more realistic model
for the dynamical evolution of the halo prior to SF in the disc
would always result in {\em less\/} dilution of the disc metallicity by 
radial flows, thus resulting in an even smaller shift in the confidence 
curves. Hence, it seems that a value of $\beta$ close to, or very slightly
larger than, 0.4 is the most favoured in order to achieve a consistent
picture with the age predictions of white dwarfs and globular clusters.

\section{Conclusions}
In order to place constraints on the most sensitive parameter of our chemical
evolution models, we have compared
in this paper the abundance ratios of heavy actinide pairs predicted by
the threshhold galactic disc models of CPT and CT with
their currently observed abundance ratios in the Solar System. We have 
assumed that these nuclei are produced in direct 
proportion to the rate of star formation and calculated their abundances 
as a function of time, allowing for their decay. The values of the abundance 
ratios $^{235}$U/$^{238}$U and $^{232}$Th/$^{238}$U at the time of formation 
of the solar system are well known and the aim of our calculations has been to 
discover at what disc age the observed ratios are attained. We have studied a 
series of disc models with all parameters fixed except one: the
parameter $\beta$ which 
determines the rate of infall of matter to the disc. We also carried out
some calculations varying the initial metallicity of the disc, produced by 
earlier stars in the halo, but found that this factor had
negligible influence on the ages predicted by our models. Thus it seems that
-- at least in the context of our models -- the heavy element abundances in
the Galactic disc are determined essentially by the stellar component of the
disc itself, and are independent of the history of the halo.

In view of other evidence on the age of the oldest nearby stars, we have
required that star formation began in the Solar neighbourhood at least
9 Gyr ago, and have incorporated this constraint into a statistical
analysis designed to determine our maximum likelihood
estimates $\beta$ as a function of the current age 
of the galactic disc, and vice versa.

Our model predictions showed that the deduced age of the Universe is a 
monotonically decreasing function of $\beta$. A high value of
$\beta$ implies both rapid 
disc formation and a strong initial burst of star formation. A value of 
$\beta \simeq 0.4$ predicts the current age
of the galactic disc to be $\sim 13.5$ Gyr, which is consistent with
current estimates of globular cluster ages, and is also in agreement 
with the models considered in CPT which satisfy other constraints on the 
properties of the local galactic disc -- such as the G-dwarf distribution, 
the age metallicity relation and the fraction of gas content.
Lower values of $\beta$ require a very large disc age to satisfy the 
cosmochronological contraints.

We have also studied the effect on the age predictions for the galactic disc of
incorporating radial flows in our chemical evolution models, following the
treatment presented in CT. We simplified the analysis by assuming that the
initial enrichment was reduced to zero everywhere in the disc in our models,
We found that, even in this limiting case, the 67\% confidence limits for 
$\beta$ at a given disc age were increased by only $\sim 0.05$; we argued that
any realistic, self-consistent model for the dynamical evolution of the halo
and the disc under radial flows would result in {\em less\/} dilution of the 
disc metallicity, and thus result in a smaller positive shift in the confidence
intervals for $\beta$.

We have also considered the consistency of the age determined by 
cosmochronology with other constraints on the age of the galaxy.
The estimated disc age for $\beta = 0.4$ is larger than that deduced 
from white dwarf cooling, but within the uncertainties of that age. Our
prediction is in any event perfectly {\em consistent\/} 
with the ages of white dwarfs 
in the Solar neighbourhood since, as pointed out by Pitts and Tayler (1992),
a delay in the onset of star formation means that the age deduced locally
from white dwarfs is somewhat less than the true disc age.

This leaves as the only major problem a reconciliation of the disc age with 
the age of the Universe obtained from current estimates of
cosmological parameters. Of course this problem is essentially no
different to that of reconciling globular cluster ages with cosmology. 
If the value of the Hubble constant is indeed greater than 
$60 {\rm kms^{-1} Mpc^{-1} }$, as recent observations suggest,
then to achieve consistency between the age of the universe and {\em both\/}
our predicted age of the disc and the age of globular clusters seems
to require a non zero cosmological constant. If globular cluster ages 
could be reduced to as little as 11 Gyr -- which has been proposed by
Chaboyer (1995) as a very robust lower limit, and which would again make
viable a cosmological model with $\rm H_0 = 60$ and $\Lambda_0 = 0$ -- 
then a higher value of $\beta$ might be more appropriate for our
chemical evolution models. A value of $\beta = 1$ can give a local disc age 
comparable to that of white dwarfs -- indicating that star formation began
in the Solar neighbourhood within 1 Gyr of the formation of the disc --
and an age of the Universe of 11 - 12 Gyr. Such a model 
would imply the presence of luminous discs at much higher redshift than in the 
case of $\beta = 0.4$ and this would therefore provide a suitable
observational test to discriminate between these two cases. We will investigate
high $\beta$ models, together with a more detailed study of the current
status of cosmological parameters, in a forthcoming paper.
\vspace{2mm}

\noindent
{\sc{acknowledgements}}

The authors would like to thank Roger Tayler for many helpful discussions,
and acknowledge the use of the STARLINK computer facilities at the
University of Sussex. MAH acknowledges the hospitality of SISSA, where
much of this work was carried out. MAH was supported by a PPARC personal
research fellowship. KC gratefully acknowledges the award of an IAU
travel grant in connection with this research.

\bibliographystyle{mn}

\begin{thebibliography}{}
\bibitem[] *Anders, E., Grevesse N., 1989, {\em Geochim Cosmochim. Acta}, 
{\bf 53}, 197
\bibitem[] *Burkert A., 1993, in `Nuclear Astrophysics: Proceedings of the
Caltech Centennial Year -- Nuclear physics symposium in honour of
William A. Fowler's 80th birthday', eds. Schramm D.N., Woosley S.E.,
{\em Phys Rep}
\bibitem[] *Butcher H.R., 1987, {\em Nature}, {\bf 328}, 127
\bibitem[] *Chaboyer B., 1995, {\em ApJ (Lett)}, {\bf 444}, L9
\bibitem[] *Chamcham K., Pitts E., Tayler R.J., 1993, {\em MNRAS}, {\bf 263}, 
967
\bibitem[] *Chamcham K., Tayler R.J., 1994, {\em MNRAS}, {\bf 266}, 282
\bibitem[] *Chiosi C., Bertelli G., Bressan A., 1992, {\em ARA\&A}, {\bf 30}, 
235
\bibitem[] *Clayton D.D., 1988, {\em MNRAS}, {\bf 234}, 1
\bibitem[] *Cowan J.L., Thielemann E.K., Truran J.W, 1991, {\em ARA \& A}, 
{\bf 29}, 447
\bibitem[] *D' Antona F., Mazzitelli I., 1990, {\em ARA \& A}, {\bf 28}, 139
\bibitem[] *Freedman W., Madore B.F., Mould J.R., Hill R., Ferrarese L.,
Kennicutt Jr. R.C., Saha A., Stetson P.B., Graham J.A., Ford H.,
Hoessel J.G., Huchra J., Hughes S.M., Illingworth G., 1994, {\em Nature}, 
{\bf 371}, 757 
\bibitem[] *Gilmore G., Kuijken K., Wyse R.F.G., 1989, 
{\em ARA \& A}, {\bf 27}, 555  
\bibitem[] *Hendry M.A., Simmons J.F.L., 1990, {\em A \& A}, {\bf 237}, 275
\bibitem[] *Hesser J.E., Harris W.E., VandenBerg D.A., 1987, {\em PASP},
{\bf 99}, 739
\bibitem[] *Iben I, Jr., Laughlin G., 1989, {\em ApJ}, {\bf 341}, 312
\bibitem[] *Klapdor-Kleingrothaus H.V., 1991, in `Nuclei in the Cosmos',
ed. Oberhummer H., (Springer-Verlag), p199
\bibitem[] *Liddle A.R., Lyth D.H., Roberts D., Viana P.T.P., 1995,
{\em MNRAS}, submitted
\bibitem[] *Malaney R.A., Fowler W.A., 1987, {\em MNRAS}, {\bf 237}, 67
\bibitem[] *Mathews G.J., Schramm D.N., 1993, {\em ApJ}, {\bf 404}, 468
\bibitem[] *Mestel L., 1952, {\em MNRAS}, {\bf 112}, 583 \& 599
\bibitem[] *Mestel L., Ruderman M.A., 1967, {\em MNRAS}, {\bf 136}, 27
\bibitem[] *Pagel B.E.J., 1989, in  
`Evolutionary Phenomena in Galaxies', eds. Beckman J.E., Pagel B.E.J.,
Cambridge University Press, p.201
\bibitem[] *Pagel B.E.J., 1993, in `Origin and Evolution of the Elements', 
eds. Prantzsos et al., Cambridge University Press, p. 465
\bibitem[] *Pitts E., Tayler R.J., 1992, {\em MNRAS}, {\bf 255}, 557
\bibitem[] *Riess A.G., Press W.H., Kirschner R.P., 1995,
{\em ApJ (Lett)}, {\bf 438}, L17
\bibitem[] *Renzini A., 1993, in `Relativistic Astrophysics and Particle 
Cosmology',  eds Akerlof C.W. and Srednicki M.A.,
Annals of the New York Academy of Sciences, Volume 688, 124
\bibitem[] *Schatzman E., 1958, `White Dwarfs', (North-Holland P.C., Amsterdam) 
\bibitem[] *Schmidt M., 1959, {\em ApJ}, {\bf 129}, 243 
\bibitem[] *Schmidt B, Kirschner R.P., Eastman R.G. , 1994, {\em ApJ}, 
{\bf 432}, 42
\bibitem[] *Schramm D.N., Wasserburg G.T., 1970, {\em ApJ}, {\bf 162}, 57
\bibitem[] *Tayler R.J., 1986, {\em Q. Jl. R. astr. Soc.}, {\bf 27}, 367
\bibitem[] *Tinsley B.M., 1980, {\em Fund. Cosm. Phys.}, {\bf 5}, 287	
\bibitem[] *VandenBerg D.A., 1990, {\em Astr. J.}, {\bf 100}, 445
\bibitem[] *van den Bergh S., 1994, {\em PASP}, {\bf 106}, 1113
\bibitem[] *van den Bergh S., 1995, {\em JRASC}, {\bf 89}, 6
\bibitem[] *Winget D.E., Hansen C.J., Liebert J., Van Horn H.M., 
Fontaine J., Nather R.E., Kepler S.D., Lamb D.Q.,
1987, {\em ApJ}, {\bf 215}, 177
\end{thebibliography}

\clearpage
\newpage
\begin{table*}
  \centering
  \caption{The range of age predictions reported in the literature for the 
four methods of age determination considered in this paper}
\begin{tabular}{llcl}
Constraint &  Method  & Age, $T_{0}$, (Gyr) & References\\ \hline
age of the solar neighbourhood & cooling of white dwarfs
& 9.3 $\pm$ 2 & Winget et al. (1987)\\
age of the galactic disc & cosmochronology
&10 - 30&Cowan, Thielemann \& Truran (1991)\\
age of the galaxy & globular clusters
& 13 - 15 $\pm$ 3  & Renzini (1993), Chaboyer (1995)\\
age of the universe & cosmological parameters &
8 - 30 & various, cf. van den Bergh (1994, 1995) \\
\end{tabular}
\end{table*}
\clearpage
\newpage
{\bf{\underline{LIST OF FIGURES}}}

\begin{itemize}
\item{1. Growth of the stellar disc radius with time: i.e. the time, 
$t_{\star}$, of the onset of SF at radius $R_{\star}$}
\item{2. Time variation of SF for different values of $\beta$}
\item{3. Time variation of the gas fraction, $\mu$,  for different values
of $\beta$. The observed value of $\mu$ and its $1 \sigma$ error band are
indicated by the horizontal lines. The upper scale is the corresponding
age of the universe, assuming an estimate of 1.5 Gyr for the
`gestation period' of the disc.}
\item{4. Time evolution of the $^{232}$Th/$^{238}$U (a) and
$^{235}$U/$^{238}$U (b) abundance ratios for different values 
of $\beta$. The observed abundance ratios and their $1 \sigma$ error bands
are indicated by the horizontal lines. The lower scale is the age of the      
disc at the time of formation of the
Solar System and the upper scale is the estimated age of the universe,
inferred from the age of the disc by adding 6 Gyr to account for the age
of the Solar System and the `gestation period' of the disc. }
\item{5. The likelihood distribution of the age of the disc
at the time of formation of solar system, conditional upon the values of 
$\beta$ assuming a minimum duration of nucleosynthesis of 
$t_{\rm{nuc}} = 4.5$ Gyr.}
\item{6. The likelihood distribution of $\beta$ conditional upon
the age of the disc at the time of formation of the solar system, and
assuming a minimum duration of nucleosynthesis of $t_{\rm{nuc}} = 4.5$ Gyr.}
\item{7. The maximum likelihood estimator of $\beta$ as a function of the
age of disc at the time of formation of the solar system, and assuming a  
minimum duration of nucleosynthesis of $t_{\rm{nuc}} = 4.5$ Gyr. }
\item{8. Same as figure 8, but now with a minimum duration of
nucleosynthesis of $t_{\rm{nuc}} = 6.0$ Gyr. }
\item{9. Confidence curves of the parameter $\beta$ as a function of the
age of the disc at the time of formation of the solar system, shown at the  
67\% (a) and 90\% (b) level, and for a minimum duration of nucleosynthesis
of 4.5 Gyr.}
\item{10. Same as figures 9a and 9b, but now with a minimum duration of
nucleosynthesis of 4.5 Gyr.}
\item{11. 67\% confidence curves for the infall parameter $\beta$ and the
age of the disc at the time of formation of the solar system, for a minimum
duration of nucleosynthesis of 4.5 Gyr and with $z_{i0} = 0$. Comparison
with figure 9a illustrates that the dependence of our results on the
adopted value of $z_{i0}$ is negligible.}
\item{12. 67\% confidence curves for the infall parameter $\beta$ and the
age of the disc at the time of formation of the solar system computed      
from our radial flow model. We assume a minimum duration of
nucleosynthesis of 4.5 Gyr and adopt the limiting value of $z_{i0} = 0$, as
explained in the text.}
\end{itemize}
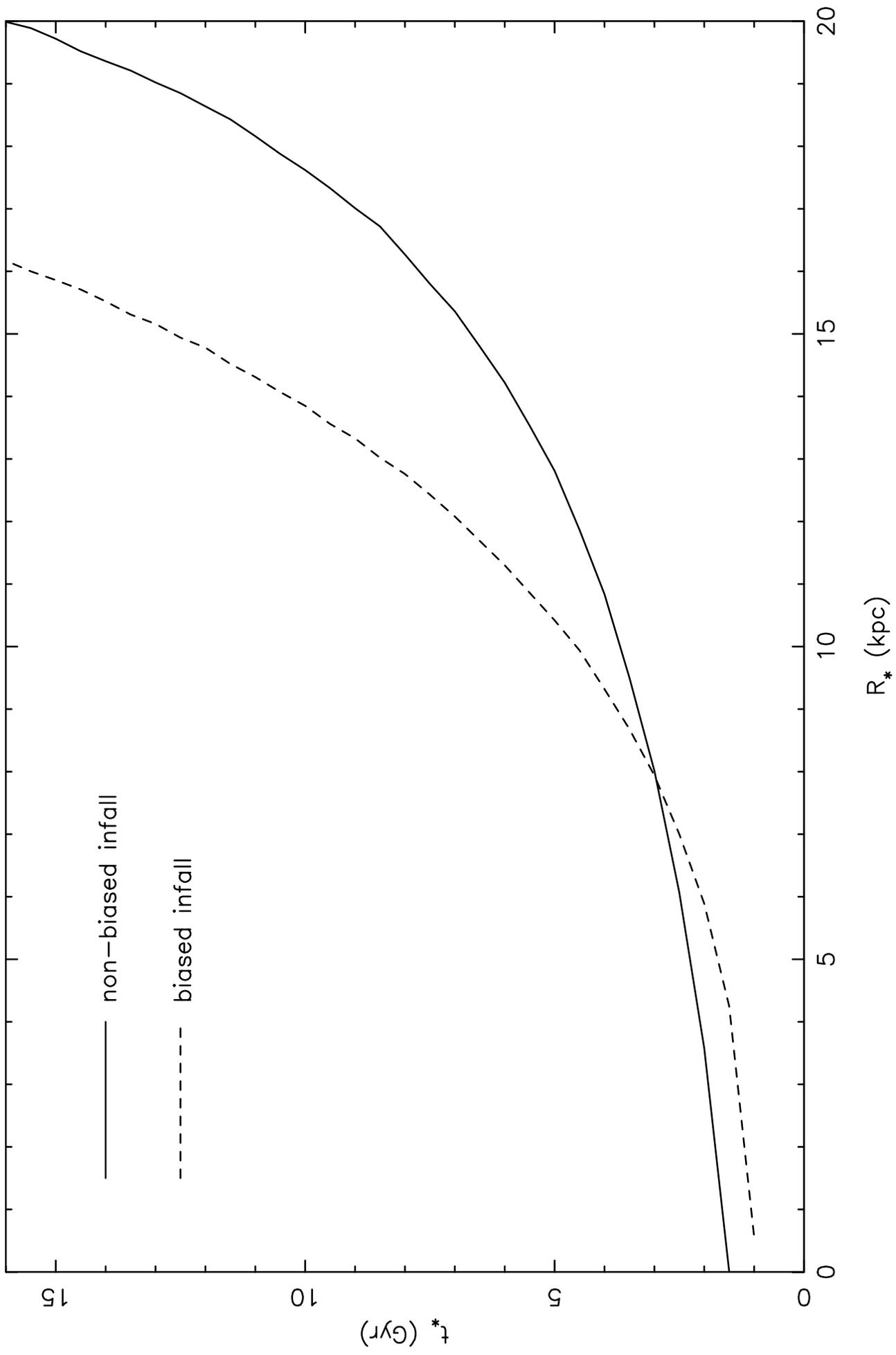
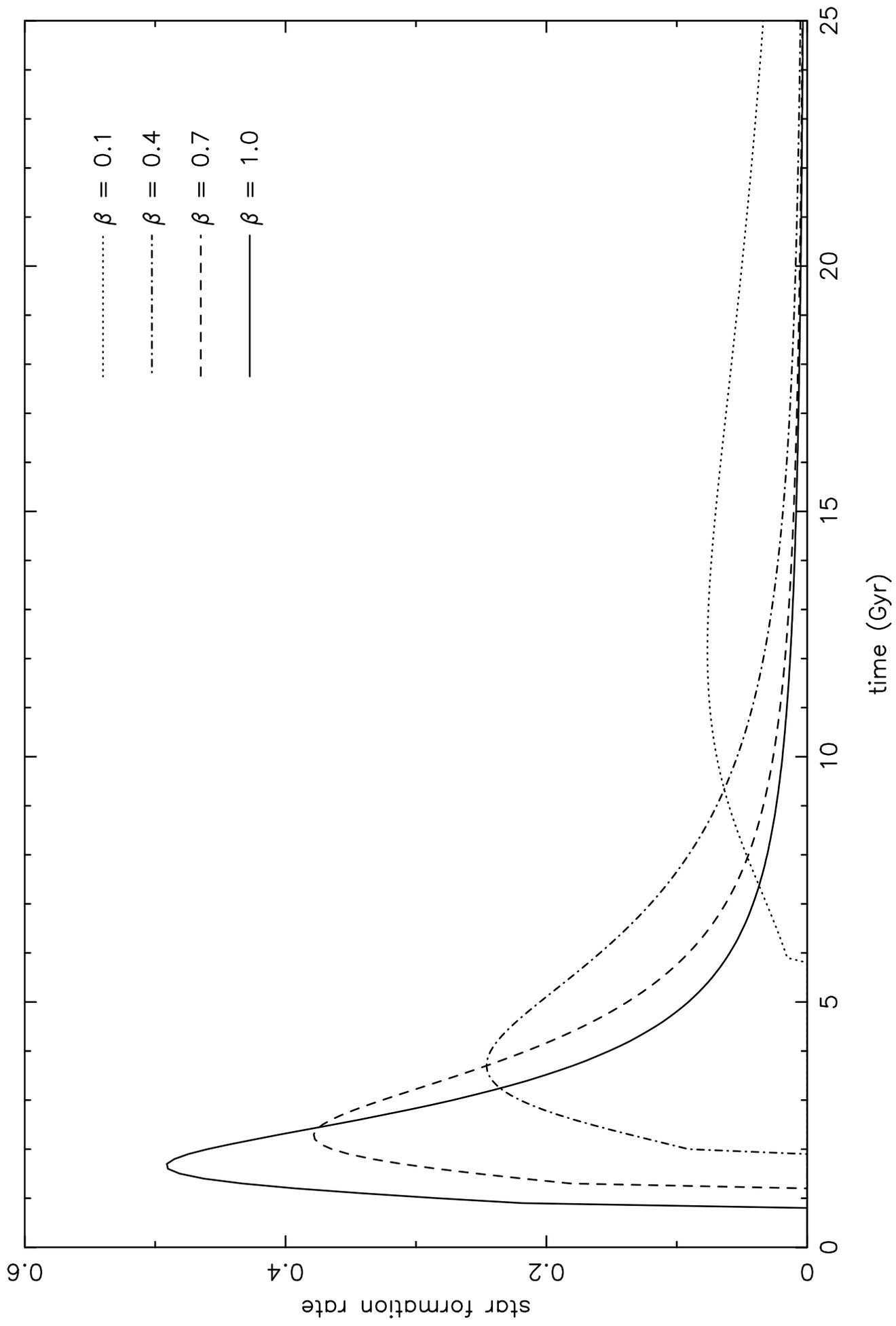
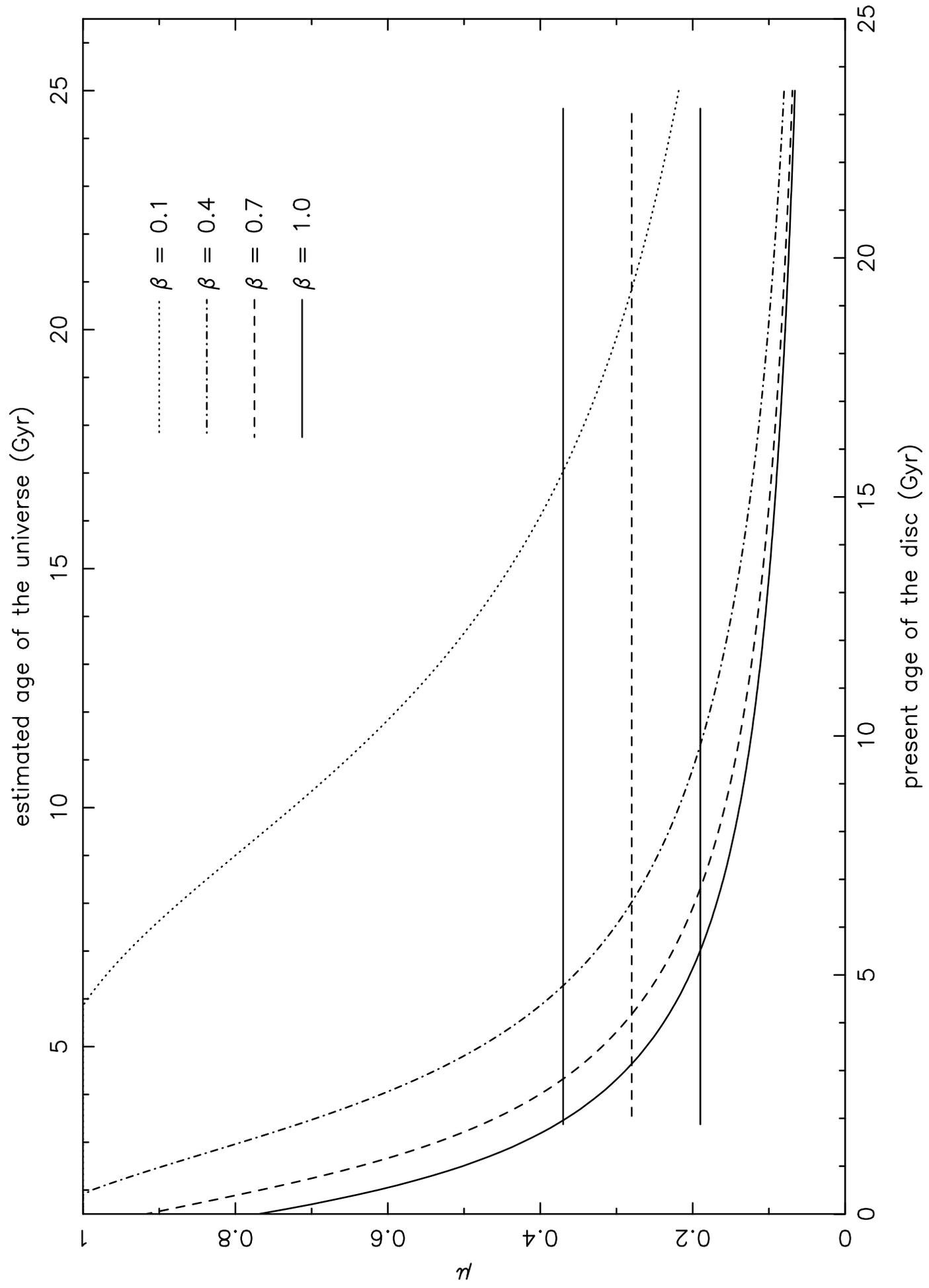
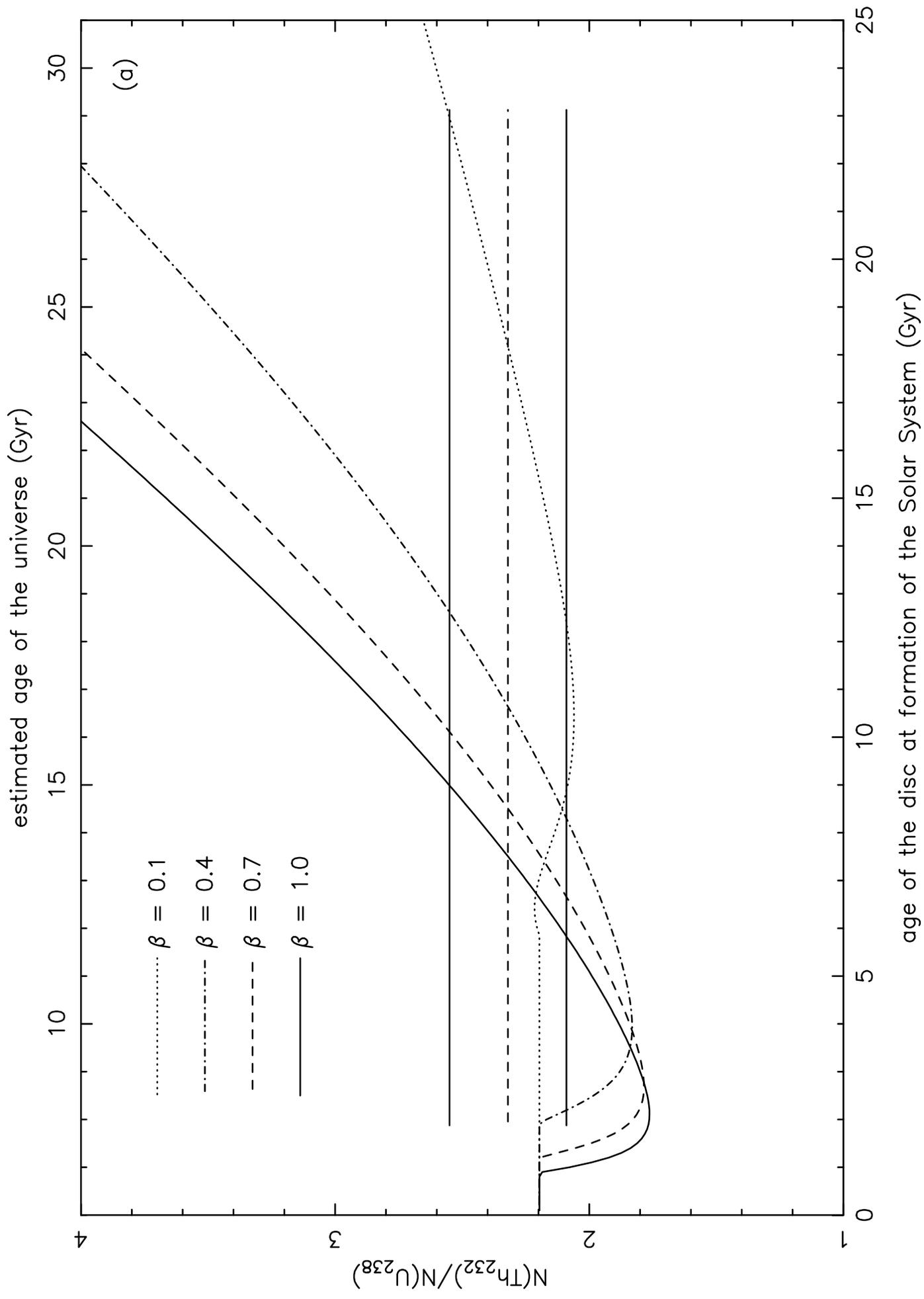
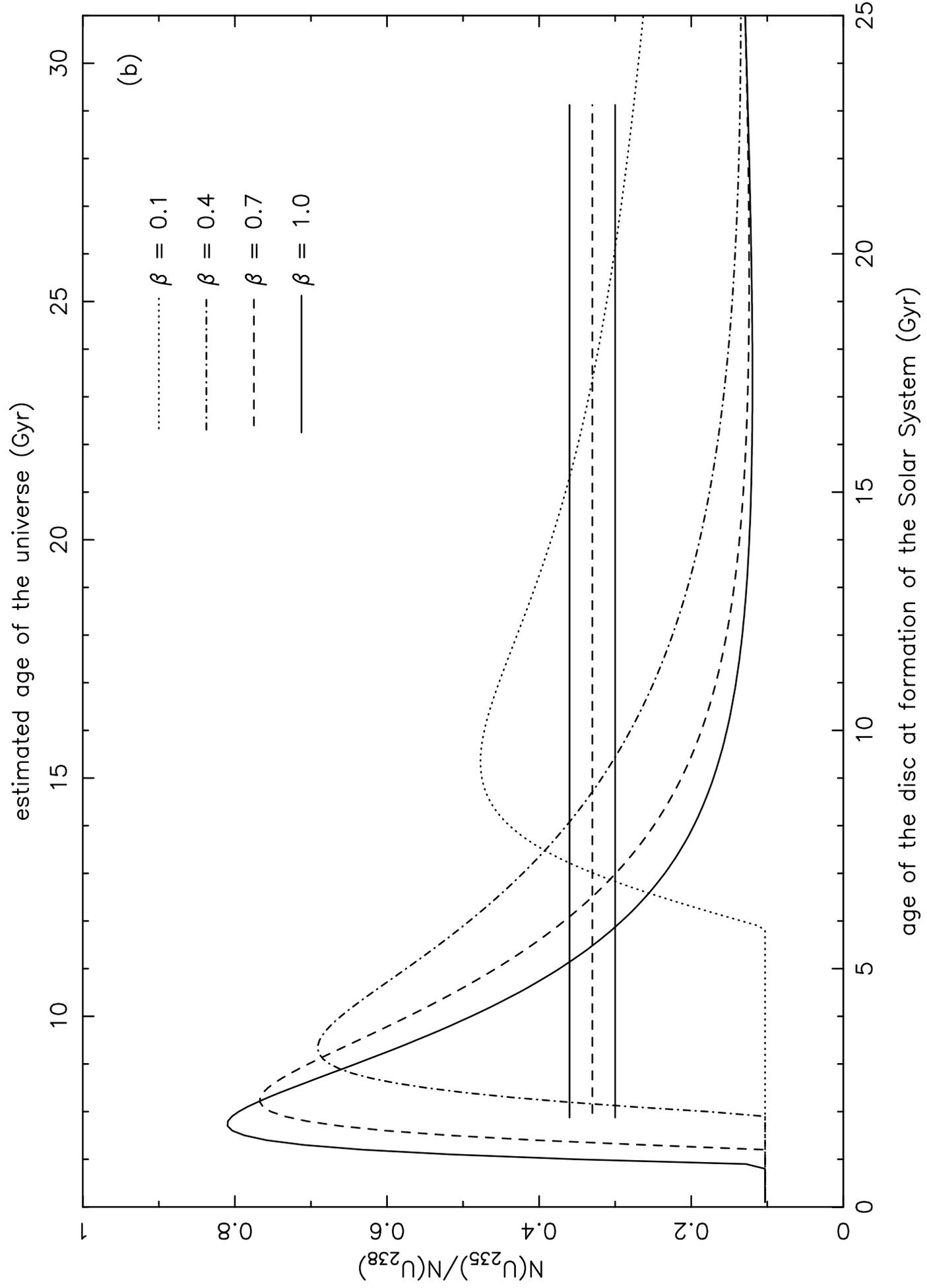
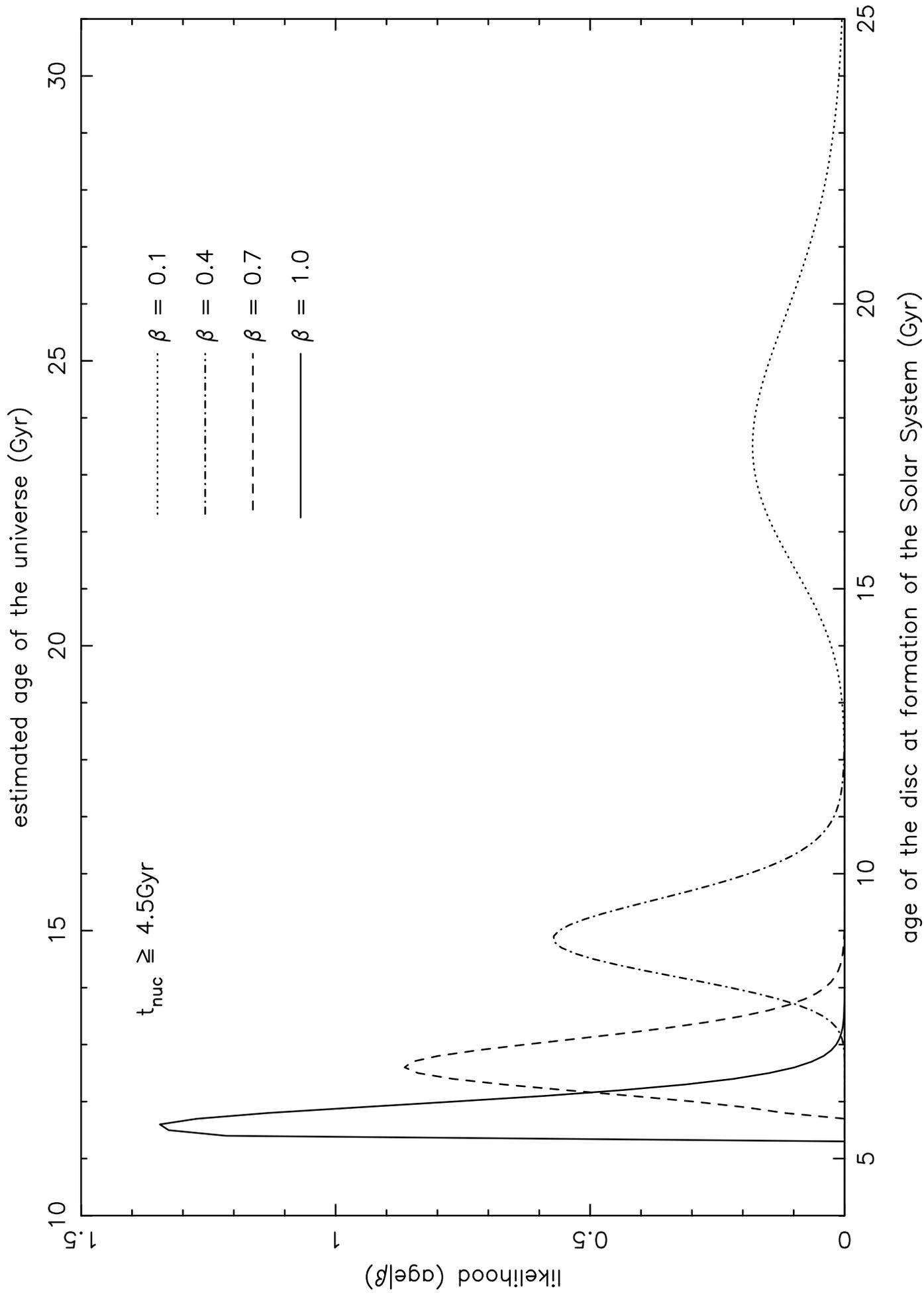
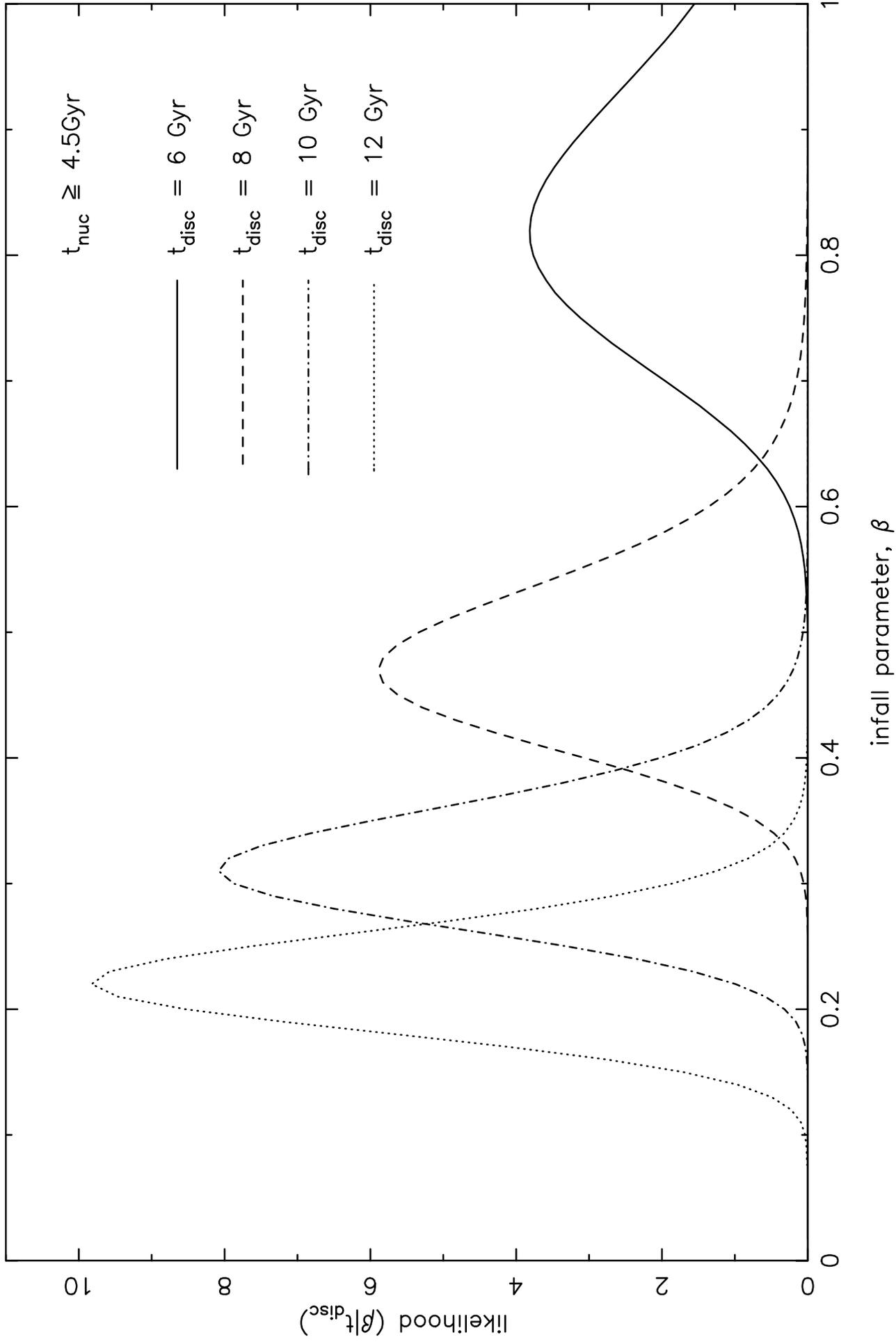
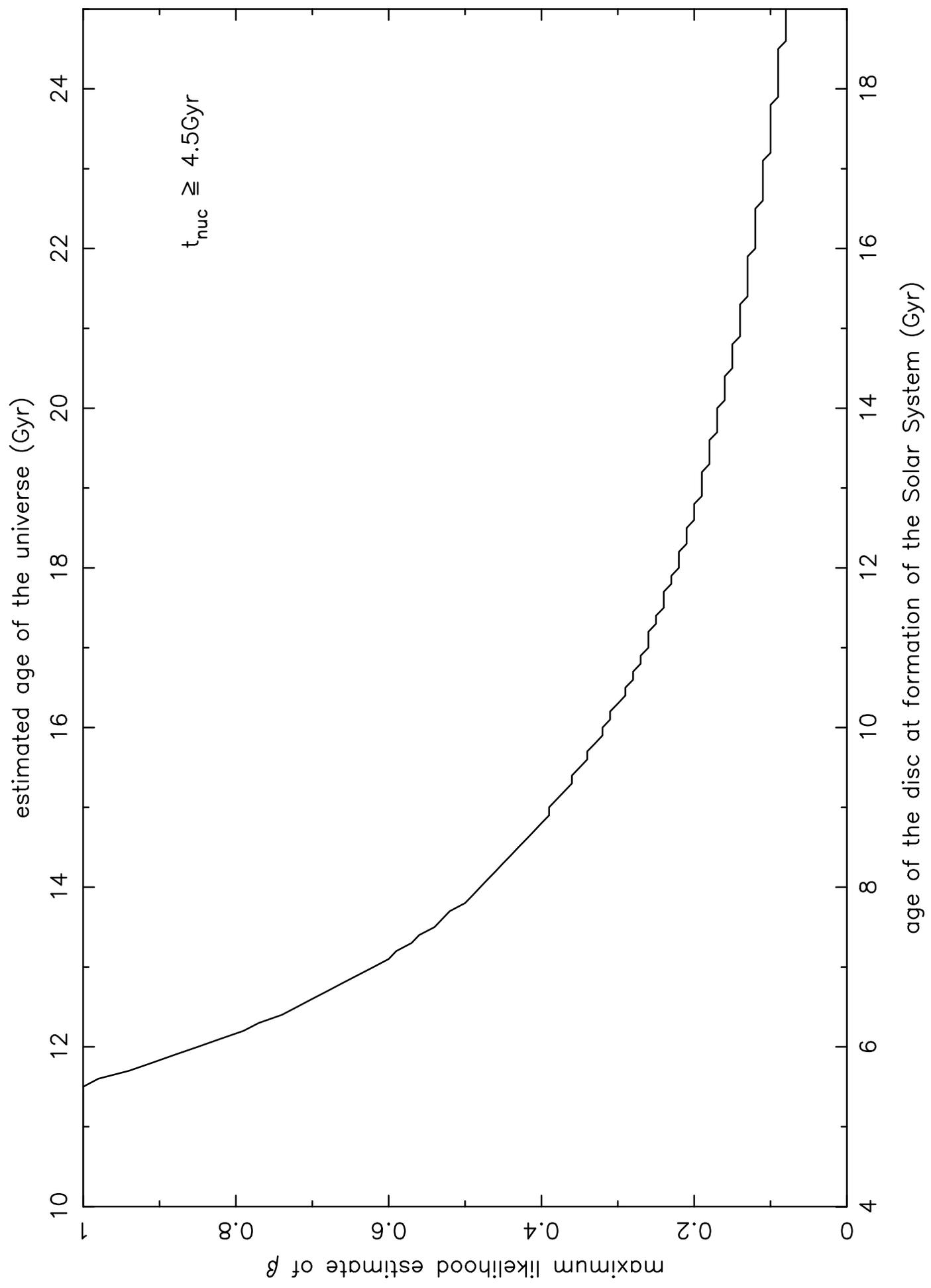
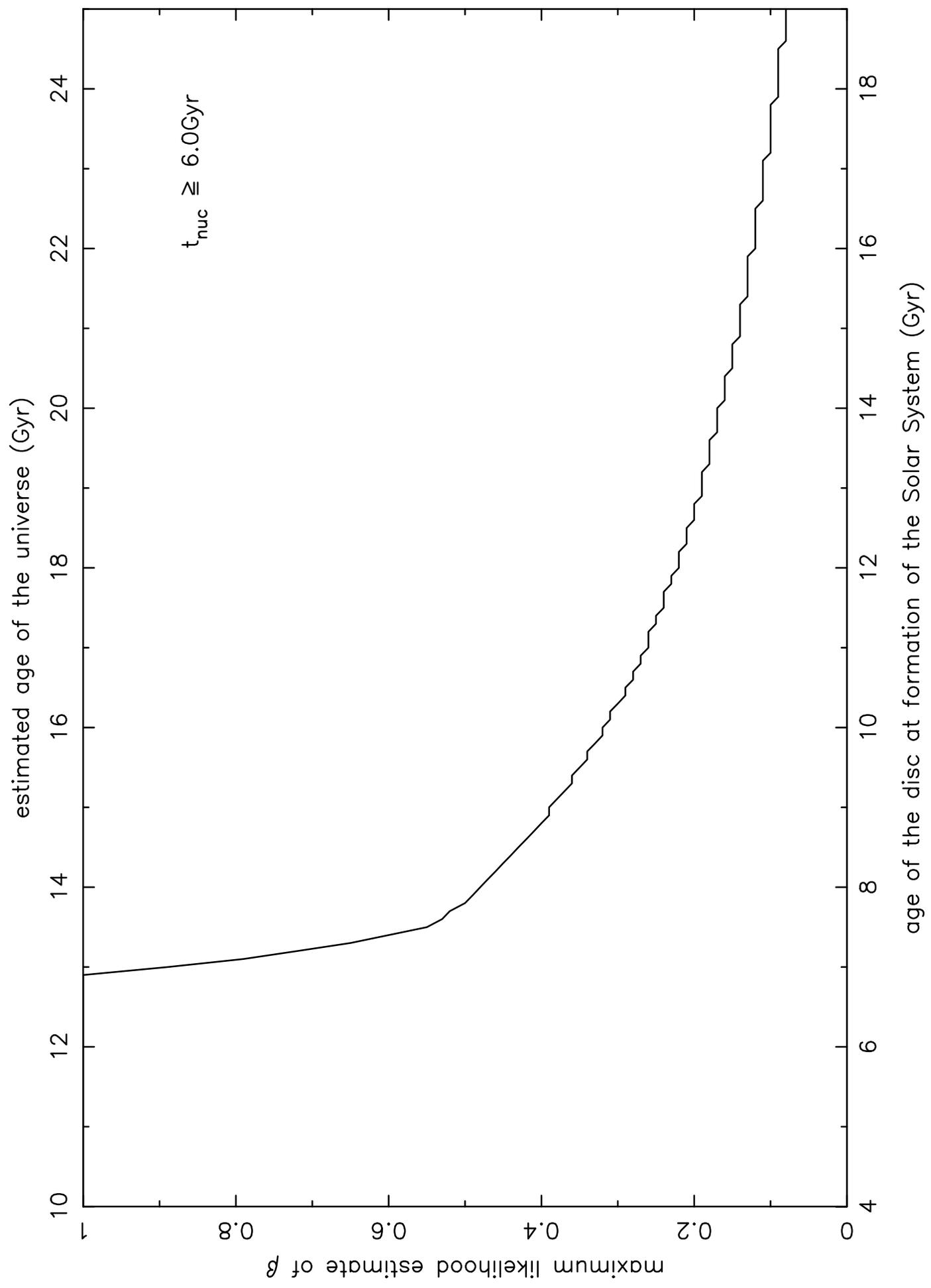
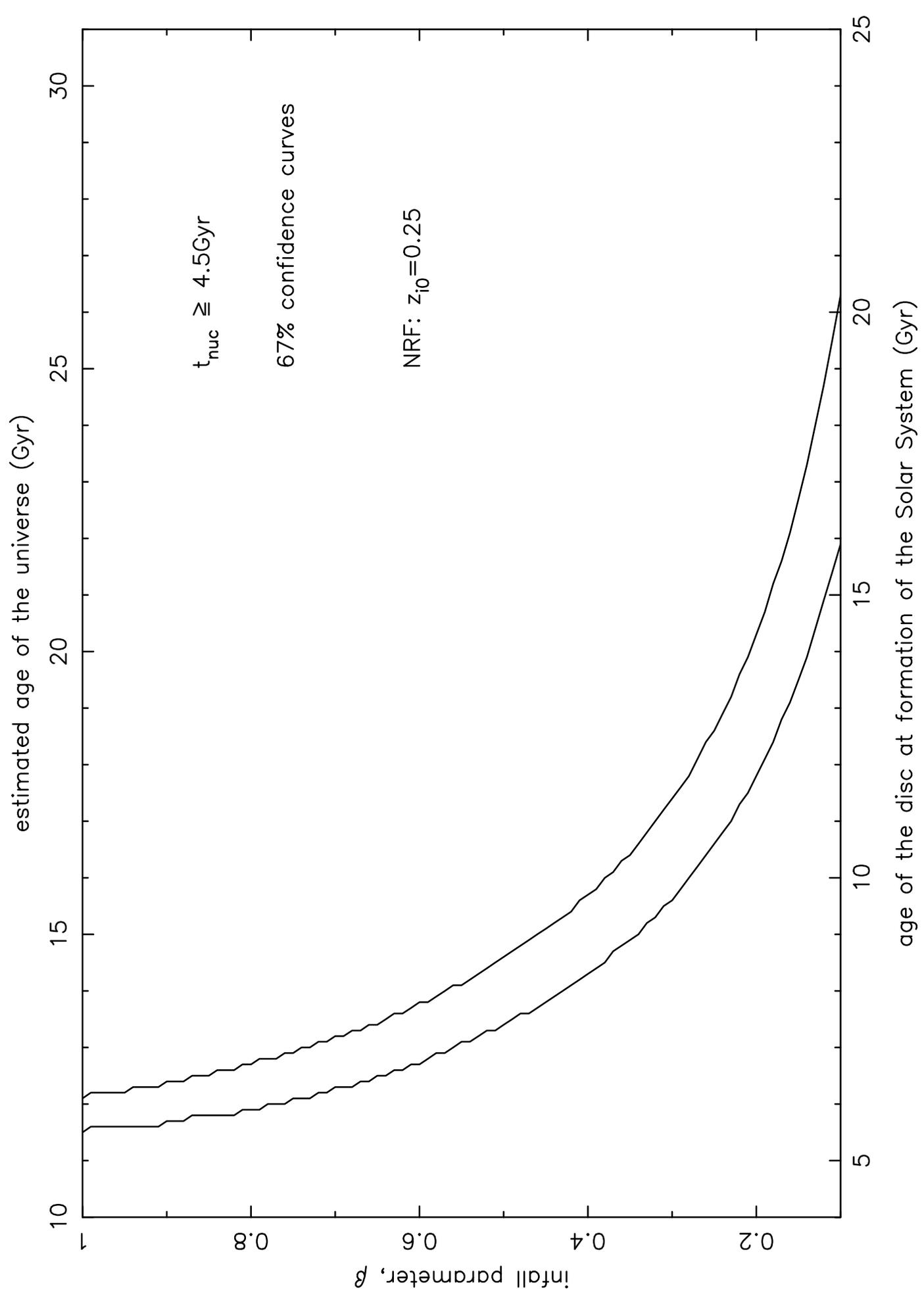
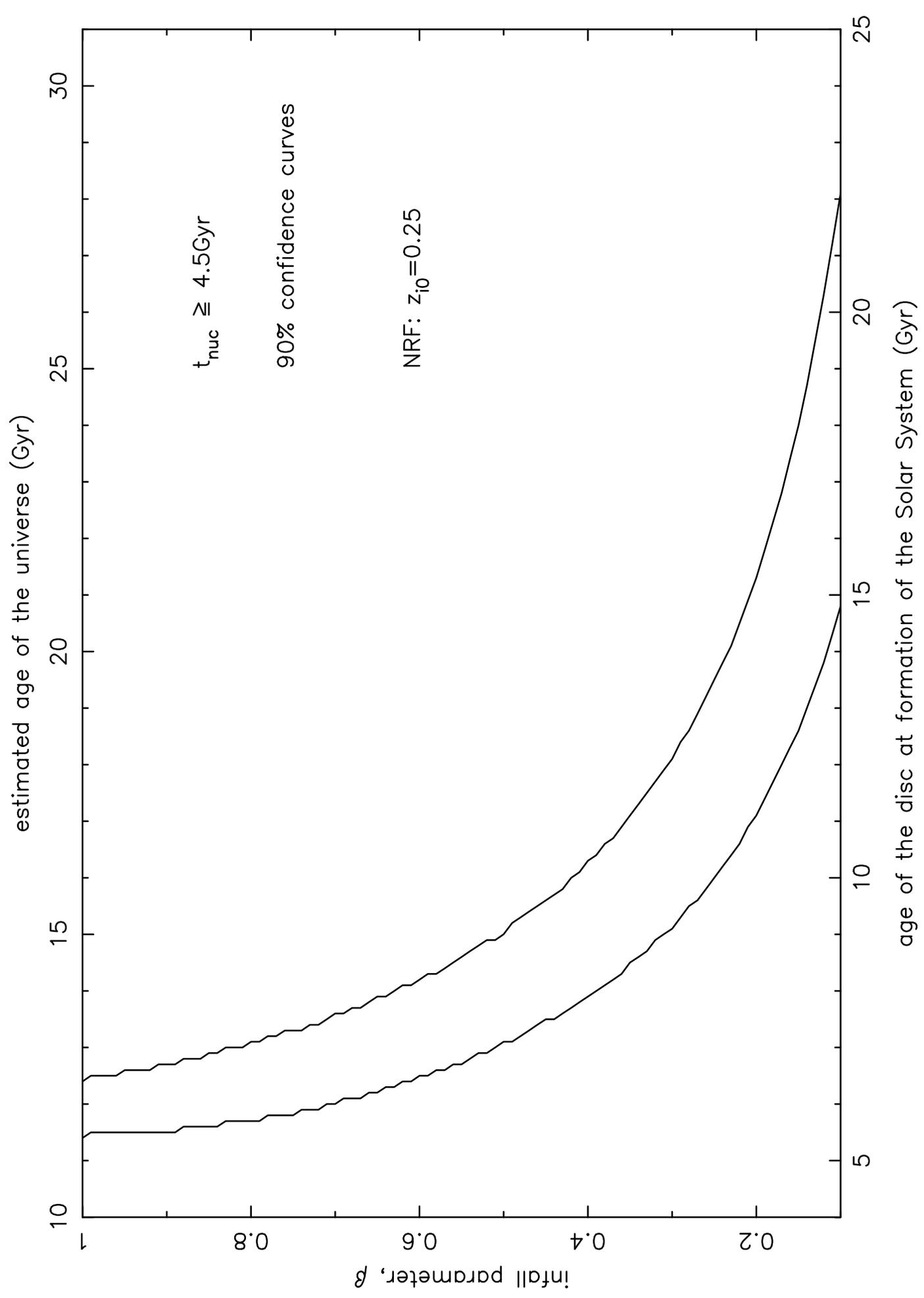
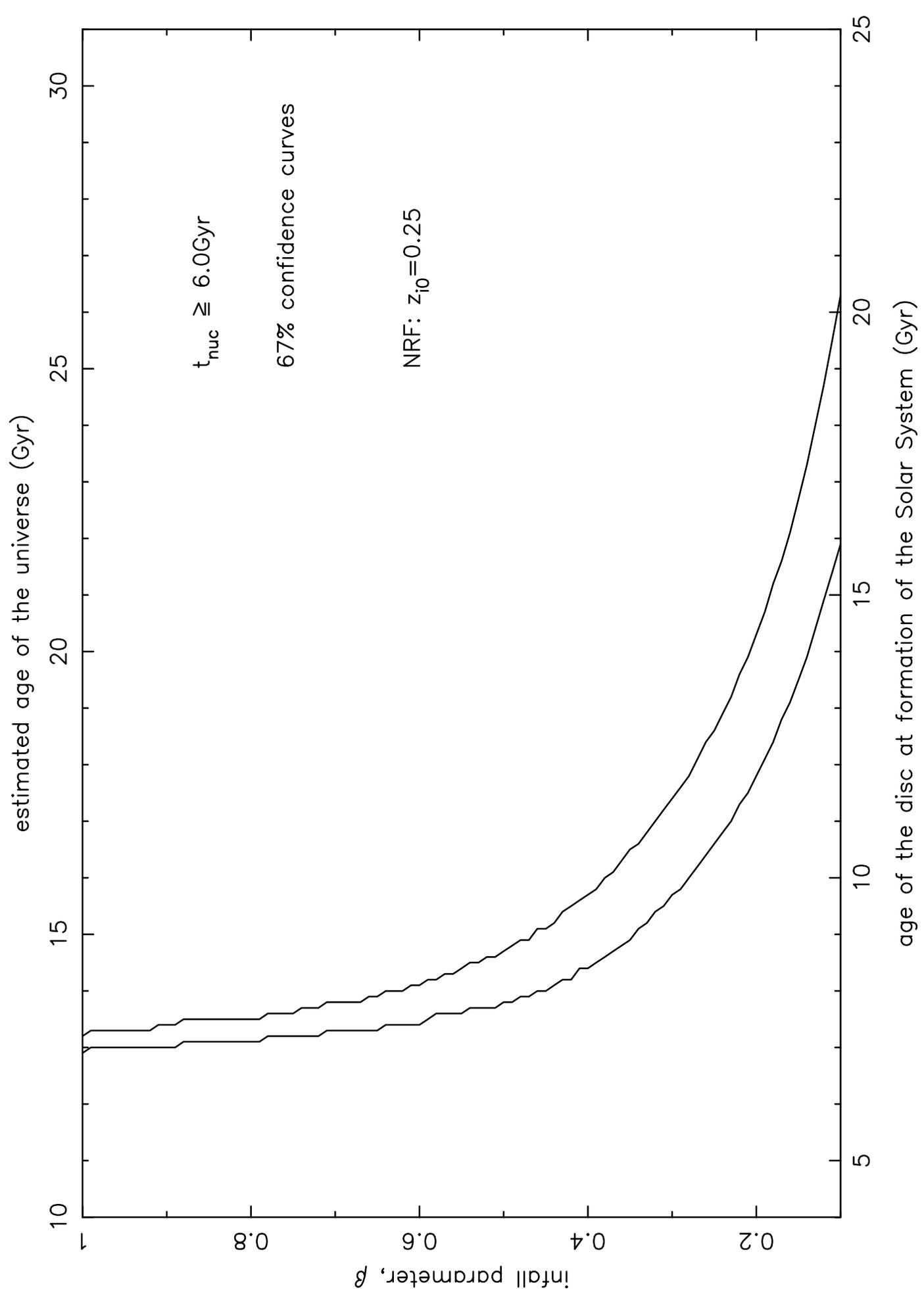
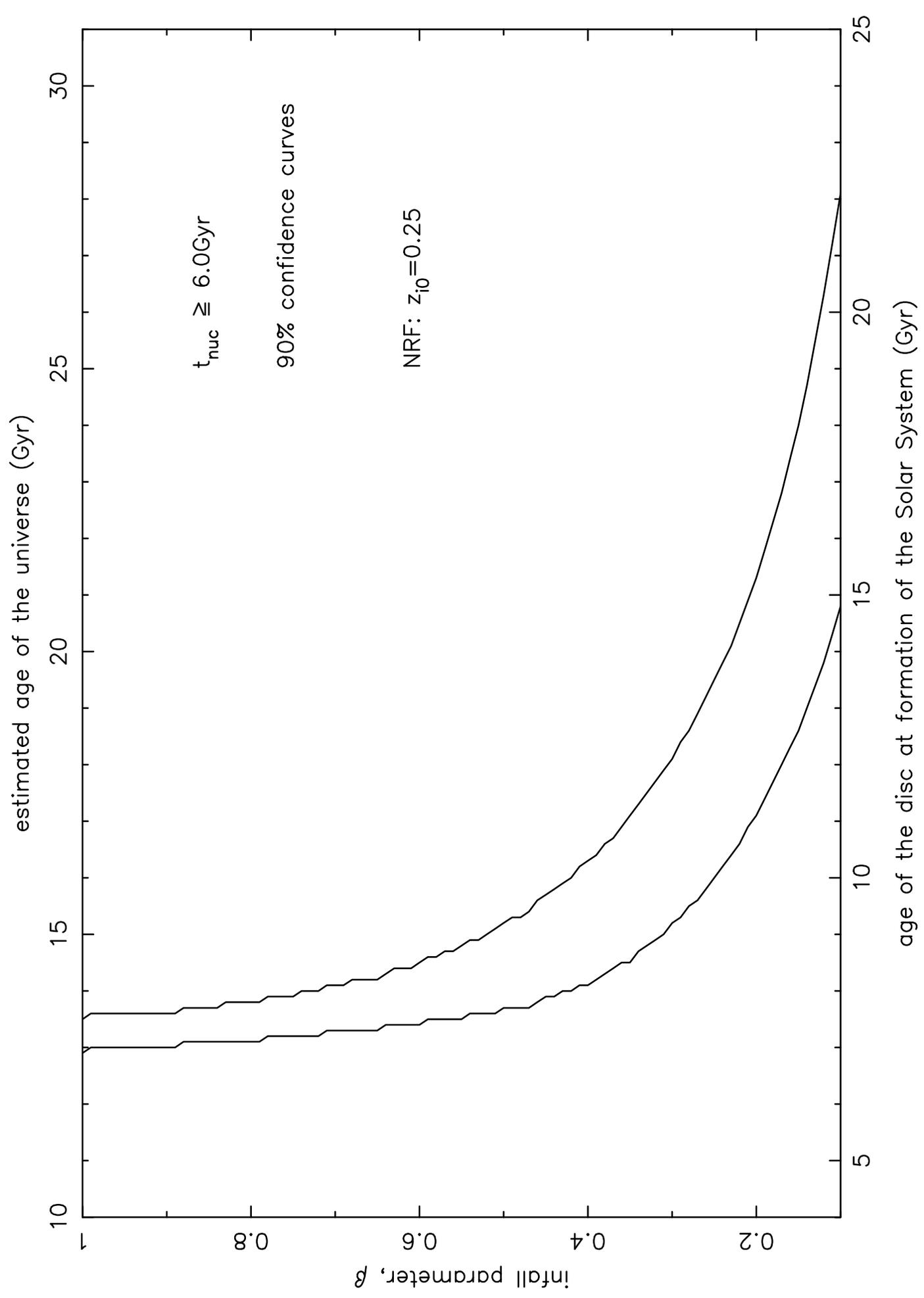
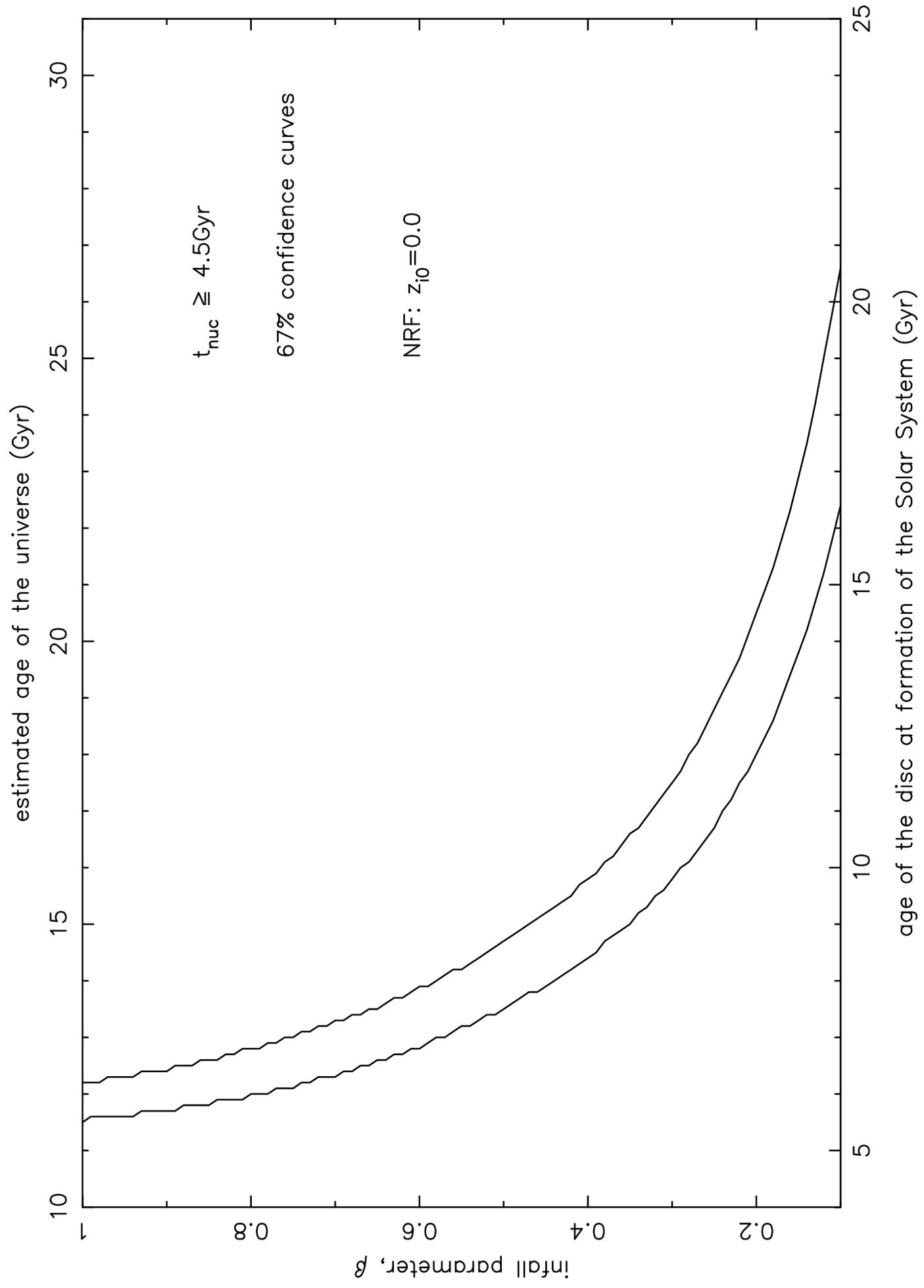
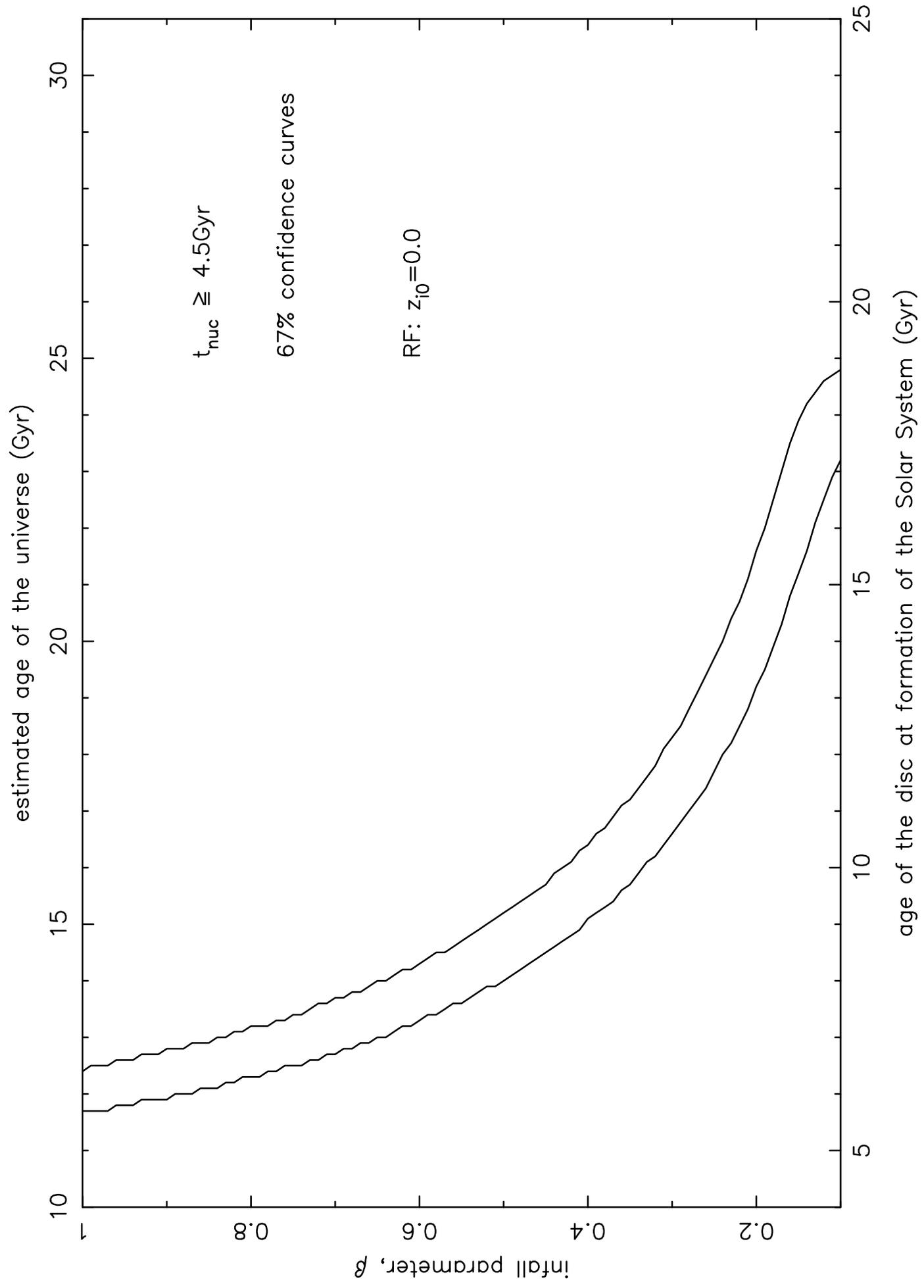
\end{document}